\documentclass[twocolumn,prb,aps,amsmath,amssymb,superscriptaddress,showpacs]{revtex4-1}
\usepackage{natbib}
\usepackage{epsfig}
\usepackage{float}
\usepackage{amsmath}
\usepackage{xcolor,colortbl}
\usepackage[utf8]{inputenc}
\usepackage[T1]{fontenc}
\usepackage{lmodern} % load a font with all the characters

\definecolor{Gray}{rgb}{0.72,0.72,0.98}
\definecolor{LightCyan1}{rgb}{0.83,0.83,0.98}
\definecolor{LightCyan2}{rgb}{0.91,0.92,1}

\renewcommand{\[}{\left[}

\begin{document}

\newcommand{\cnr} {Istituto di Struttura della Materia of the National Research Council, Via Salaria Km 29.3,I-00016 Monterotondo Stazione, Italy}
\newcommand{\etsf} {European Theoretical Spectroscopy Facilities (ETSF)}
\newcommand{\spaulo} {Instituto de F\'{\i}sica Te\'{o}rica, Universidade Estadual Paulista (UNESP), Rua Dr. Bento T. Ferraz, 271, S\~{a}o Paulo, SP 01140-070, Brazil.}

\title{Electron-phonon scattering effects on electronic and optical properties of orthorhombic GeS}

\author{Cesar E. P. Villegas}
\affiliation{\cnr}
\affiliation{\spaulo}

\author{A. R. Rocha}
\affiliation{\spaulo}

\author{Andrea Marini}
\affiliation{\cnr}
\affiliation{\etsf}

%\date{\today}
\begin{abstract}
Group-VI monochalcogenides are attracting a great deal of attention due to their peculiar
anisotropic properties. Very recently, it has been suggested that GeS could act as a promissory absorbing material with 
high input-output ratios, relevant features for designing prospective optoelectronic devices. In this work, we
use the \emph{ab-initio} many body perturbation theory to study the role of the electron-phonon coupling on orthorhombic GeS. 
We identify the vibrational modes that efficiently couple with the electronic states responsible for giving rise to 
the first and second excitonic state. We also study the finite-temperature optical absorption and
show that even at $T\to0K$, the role of the electron-phonon interaction is crucial to properly describe the main experimental
excitation peaks position and width. Our results suggest that the electron-phonon coupling is essential to properly describe
the optical properties of the monochalcogenides family.

\end{abstract} 
%\pacs{ }
%
% 71.38.-k 	Polarons and electron-phonon interactions
% 63.20.dk 	First-principles theory
% 79.60.Fr	Photoemission and photoelectron spectra. Polymers; organic compounds 
% 78.20.-e 	Optical properties of bulk materials and thin films
%
% 73.21.-b 	Electron states and collective excitations in multilayers, quantum wells, mesoscopic, and nanoscale systems
% 63.22.-m 	Phonons or vibrational states in low-dimensional structures and nanoscale materials
% 63.20.kd 	Phonon-electron interactions
%
%
\maketitle
\section{Introduction}
Layered materials have become prospective platforms for a next generation of technological devices due to 
their widely tunable physical properties.\cite{layer1,layer2} Their potential applications
include optoelectronic,\cite{opto} photovoltaic,\cite{photo,photo2} and thermoelectric\cite{thermo,thermo2} properties. Besides
the well-characterized hexagonal layered crystals such as: graphite,\cite{graph} boron nitride,\cite{bn1,bn2}
and transition metal dichalcogenides,\cite{tmdc1,tmdc2} recently, black phosphorus (BP)\cite{BPrev,BPrev2} and group-VI monochalcogenides
(GeS, GeSe, SnSe, and SnS)\cite{mono1,mono2} have gained renewed attention, fundamentally, as a direct 
consequence of their unique anisotropic properties. These layered materials possess an orthorhombic crystalline
structure (\emph{Pnma} space group) with a puckered atomic arrangement. This peculiarity gives rise to well-defined
zigzag and armchair directions that are crucial in determining a number of interesting phenomena.\cite{villegas2,villegasexci,aniso1,thermo,mono-ani,villegas1}
 
Orthorhombic GeS exhibits an optical gap of $\sim$1.65 eV\cite{GeSgap} that, together with its low toxicity\cite{toxi} and stability
under normal conditions -superior to the one observed in BP\cite{toxibp,stabp}- represents an appealing 
candidate for optoelectronic applications.
In fact, a recent report indicates that GeS-based photodetectors
could provide high ratios of external quantum efficiency and detectivity that are comparable with the ones found in current
commercial photodetectors.\cite{GeSprop1} In addition, it has been suggested that GeS could also have thermoelectric potential
due to low values of thermal conductivity.\cite{thermo2} These potential applications, highly temperature dependent, 
exhort to a comprehensive understanding of the finite-temperature properties of GeS.

While extensive experimental characterization of GeS, including optical measurements, have been carried out,\cite{GeSgap,cardonaGeS,GeSvibra,GeSarpes,GeSopt3}
\emph{ab-initio} calculations of its fundamental properties are yet scarce. 
Albeit the accurate prediction of GeS vibrational properties,\cite{GeSvibra} the prediction of its optical spectrum
still present some discrepancies.\cite{GeSoptic,mono1,GeSopt4,lda-ges} For instance, Makinistian and Albansi,\cite{GeSoptic} studied the
optical properties of GeS using different light-polarizations and pressures. Nonetheless, the authors used advanced 
methodologies based on the Bethe-Salpeter equation (BSE) combined with many-body perturbation theory (MBPT), their 
calculations fail in predicting the relative exciton intensities and peak positions. This is likely a consequence of comparing their
results with finite temperature measurements. As a common feature, other state-of-the-art optical studies on GeS\cite{mono1,GeSopt4,lda-ges} consider atoms
frozen in their equilibrium positions thus neglecting the role of the electron-phonon (EP) interaction. 
Without including the EP interaction, temperature effects
can not be described. Even at $T\to0K$ the electronic states can be strongly renormalized by the quantum nature of atoms, this is
the quantum zero-point motion (ZPM) effect.\cite{zpm,feliciano1}

From the discrepancies between experiments and current levels of theory one can conclude that the electron-phonon (EP)
coupling plays a non-trivial role in GeS. In addition, the EP coupling could also play an important role
in the entire family of monochalcogenide crystals. This is supported by recent theoretical works addressing the finite-temperature
effects of semiconductors through the inclusion of EP coupling.\cite{hiroki2014,marini2008,alejandro2016,feliciano0,ponce}

In fact, it is well-known that lattice vibrations can affect the optical properties of semiconductors leading to changes in
the exciton peak position, linewidth and selection rules.\cite{Cardonarev,cardonabook,exci-ph} Therefore, 
it is important to include temperature effects on the state-of-the-art simulations to better
describe the optical properties and to elucidate the dynamics of the EP scattering mechanisms.

In this work, we provide a theoretical description of the effect of the EP coupling on the electronic and optical
properties of orthorhombic GeS. We clearly identify the most important phonon modes that efficiently couple
with the electronic states responsible for giving rise to the first and second excitonic states.
Our results show that, at the band-edge, the infrared longitudinal $B_{2u}$ mode is the main scattering source
for the electronic states. This mode also couples efficiently with the first excitonic state. In contrast,
the electronic states giving rise to the second excitonic state couples mostly with modes $A^{1}_{g}$ and $B^{2}_{3g}$.
We also calculate the finite-temperature optical absorption and shown that even at $T\to0K$, the role of EP coupling is crucial to 
better describe the absorption linewidth and exciton peak positions.

\section{Theory and Methodology}
%%%%%%%%%%%%%FIG1%%%%%%%%%%%%%%%%%%%%%%%%%%%%%%%%%
\begin{figure}[t]
%\centering
%\includegraphics[width=1.0\columnwidth,height=6cm]{fig1}
\includegraphics[width=0.85\columnwidth]{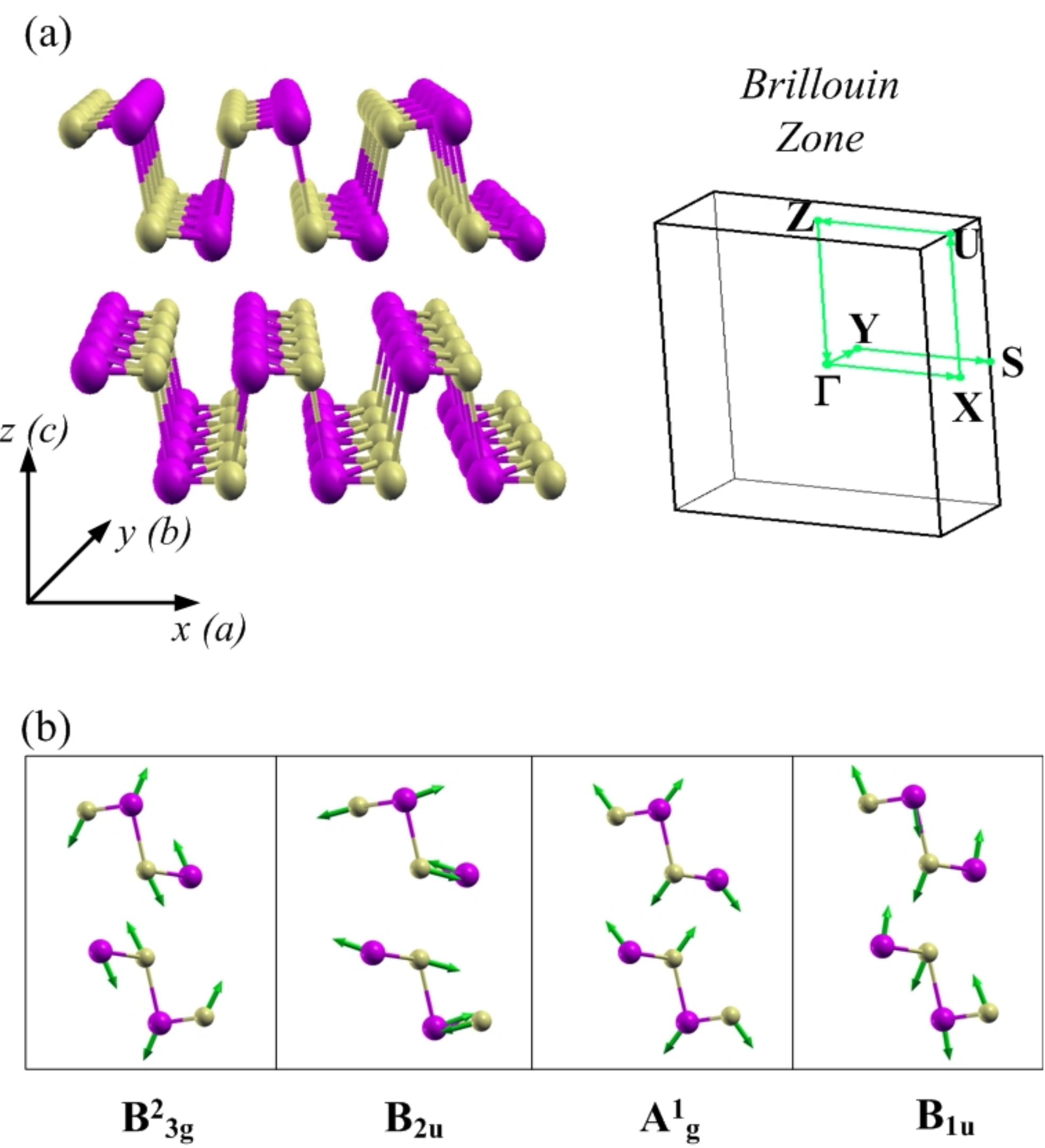}
\caption{\small{ (a) Schematic representation of the crystal structure and Brillouin zone of orthorhombic GeS showing the 
main crystallographic directions. (b) Representative atomic vibrations of GeS.
The Mulliken notation for the vibrational modes is chosen by their resemblance with black phosphorus vibrational modes.
}}\label{fig1} 
%{\rule{4cm}{2cm}}
\end{figure}
%%%%%%%%%%%%%FIG1%%%%%%%%%%%%%%%%%%%%%%%%%%%%%%%%%
We consider a pristine GeS crystal as shown in Fig. (\ref{fig1}-a). It consists of a layered puckered
structure containing 8 atoms in the orthorhombic unit cell for which each Ge atom is bonded to three S atoms. 
The obtained fully relaxed lattice parameters ($a$=4.30 \AA, $b$=3.67 \AA, and $c$=10.63 \AA) \ are in good agreement with
experimental ones.\cite{GeSvibra}

Our study is conducted in three steps. First, plane-wave density functional theory is used 
to obtain the electronic ground-state. The Perdue-Burke-Ernzerhof (PBE) exchange-correlation potential
including van der Waals corrections within the semi-empirical dispersion scheme (PBE-D) is used. We employed norm-conserving pseudopotentials,
a 60 Ry kinetic energy cutoff and a k-sampling grid in the Monkhorst-Pack scheme of 8$\times$8$\times$4 as implemented in the
Quantum-Espresso code.\cite{pwscf} The structures were fully optimized to their equilibrium position with forces smaller than 0.002 eV/\AA.

Next, we use density functional perturbation theory (DFPT)\cite{dfpt} to compute the vibrational frequencies $\omega_{\textbf{q}\lambda}$
and the derivatives of the self-consistent Khon-Sham potential with respect to the atomic displacements, needed to evaluate the 
electron-phonon coupling matrix elements. %, eqs. (\ref{eq5.1}) and (\ref{eq5.2}). 

Many-Body perturbation theory (MBPT)\cite{Onida2002} is used to describe the temperature dependent electronic 
states. There, the electron-phonon interaction is treated perturbatively\cite{marini2014,marini2015} by considering the
first and a second order Taylor expansion in the nuclear displacement, commonly known as the Fan and Debye-Waller (DB) terms,
respectively. The corresponding interacting Green's function, whose poles define the quasiparticle excitations, can be written as
\begin{equation}
G_{n\textbf{k}}(\omega,T)=[\omega-\epsilon_{n\textbf{k}}-\Sigma^{Fan}_{n\textbf{k}}(\omega,T)-\Sigma^{DW}_{n\textbf{k}}(T)]^{-1}, \label{eq1}
\end{equation}
where $\epsilon_{n\textbf{k}}$ is the Khon-Sham ground-state eigenenergies for frozen atoms.
$\Sigma^{Fan}$ is the Fan contribution 
\begin{equation}
 \begin{split}
 \Sigma^{Fan}_{n\textbf{k}}(i\omega,T)=\sum_{n'\textbf{q} \lambda}\frac{|g^{\textbf{q}\lambda}_{nn'\textbf{k}}|^2}{N}\Big[\frac{N_{
 \textbf{q}\lambda}(T)+1-f_{n'\textbf{k-q}}}{i\omega-\epsilon_{n'\textbf{k-q}}-\omega_{\textbf{q}\lambda}} \\
+\frac{N_{\textbf{q}\lambda}(T)+f_{n'\textbf{k-q}}}{i\omega-\epsilon_{n'\textbf{k-q}}+\omega_{\textbf{q}\lambda}}\Big], \label{eq4}
 \end{split}
\end{equation}
and $\Sigma^{DW}$ is the Debye-Waller term
\begin{equation}
  \Sigma^{DW}_{n\textbf{k}}(T)=-\frac{1}{2}\sum_{n'\textbf{q}\lambda}\frac{\Lambda^{\textbf{q}\lambda}_{nn'\textbf{k}}}{N}
  \Big[\frac{2N_{\textbf{q}\lambda}(T)+1}{\epsilon_{n\textbf{k}}-\epsilon_{n'\textbf{k}}}\Big]. \label{eq5}
\end{equation}
Here $N_{\textbf{q}\lambda}$ and $f_{n'\textbf{k-q}}$ represent the Bose-Einstein and Fermi-Dirac distribution functions, while
$N$ is the number of $\textbf{q}$ points in the Brillouin zone. This last $\textbf{q}$-mesh is taken randomly to better map out the phonon
transferred momentum \cite{q-random}. We include 200 electronic bands and 60 random \textbf{q}-points
for the phonon momentum to evaluate Eq. (\ref{eq4}) and Eq. (\ref{eq5}).

The electron-phonon coupling matrix elements $g^{\textbf{q}\lambda}_{nn'\textbf{k}}$, which represent
the probability amplitude for an electron to be scattered due to emission or absorption of phonons, is given by
\begin{equation}
\begin{split}
g^{\textbf{q}\lambda}_{nn'\textbf{k}}=\sum_{s \alpha}[2M_{s}\omega_{\textbf{q} \lambda}]^{-1/2}e^{i\textbf{q}.\tau_{s}} \xi_{\alpha}(\textbf{q} \lambda|s) \\
\times \langle n'\textbf{k-q}|\frac{\partial V_{scf}(\textbf{r})}{\partial R_{s \alpha}}|n\textbf{k} \rangle, \label{eq5.1}
\end{split}
\end{equation}
where, $M_{s}$ is the atomic mass of the s$-$th atom, $\tau_{s}$ is the position of the atomic displacement in the unit cell,
$\xi_{\alpha}(\textbf{q}\lambda|s)$ are the components of the phonon polarization vectors, and $V_{scf}(\textbf{r})$ is the self-consistent
DFT ionic potential. 
The second-order electron-phonon matrix elements are given by
\begin{equation}
 \begin{split}
\Lambda^{\textbf{q}\lambda,\textbf{q'}\lambda'}_{nn'\textbf{k}}=\frac{1}{2}\sum_{s}\sum_{\alpha \beta} \frac{\xi_{\alpha}^{*}(\textbf{q}\lambda|s)
\xi_{\alpha}(\textbf{q}'\lambda'|s)}{2M_{s}[\omega_{\textbf{q}\lambda}\omega_{\textbf{q}'\lambda'}]^{1/2}} \\ 
\times \langle n'\textbf{k-q-q}' |\frac{\partial^{2} V_{scf}(\textbf{r})}{\partial R_{s \alpha} \partial R_{s \beta}}|n\textbf{k} \rangle. \label{eq5.2}
\end{split}
\end{equation}
Eq. (\ref{eq5.2}) is rewritten, using translational invariance, in terms of the first order gradients only (see Ref. \onlinecite{marini2014} and \onlinecite{marini2015}).

The EP quasiparticle corrections to the Khon-Sham eigenenergies are calculated within the quasiparticle approximation (QPA).
It consists in expanding, to first-order, the self-energy frequency dependence around the bare energies. In this way, one can 
write the temperature dependent EP electronic states as\cite{Marini20091392}
\begin{equation}
E_{n\textbf{k}}(T) \approx \epsilon_{n\textbf{k}}+Z_{n\textbf{k}}(T)[\Sigma_{n\textbf{k}}^{Fan}(\epsilon_{n\textbf{k}},T)+\Sigma_{n\textbf{k}}^{DW}(T)],
 \label {eq6}
\end{equation}
where $Z_{n\textbf{k}}(T)$=$\big[1$-$\frac{\partial \Re \Sigma_{n\textbf{k}}^{Fan}(\omega)}{\partial \omega}\big|_{\omega=\epsilon_{n}}\big]^{-1}$
is the renormalization factor. Given that the Fan self-energy term is a complex function, it provides both the energy renormalization shift 
and the intrinsic quasiparticle lifetime. 

The renormalization factor can be interpreted as 
the quasiparticle charge, and, constitutes a useful tool to assess the validity of the QPA; the closer $Z_{n\textbf{k}}$ is to 1, the
more appropriate is the QPA. In fact, by assuming the validity of the QPA, one is able to rewrite Eq.(\ref{eq1}) as
\begin{equation}
G_{n\textbf{k}}=\frac{Z_{n\textbf{k}}}{\omega-E_{n\textbf{k}}(T)},
\end{equation}
whose spectral function (SF), $A_{n\textbf{k}}$=$\pi^{-1}$ $\Im{[G_{n\textbf{k}}]}$, should resemble a Lorentzian function centered at
$E_{n\textbf{k}}$. As the lattice vibrations become stronger, the SF gets wider and extends over large energy windows.
Note, however, that one should be aware of the breakdown of the QPA that can be recognized by
asymmetries and the appearance of new peaks in the SF.\cite{alejandro2016,cannu,gali}

Finally, the temperature dependent excitonic effects are included on top of the
frozen-atom Bethe-Salpeter equation (BSE)
\begin{equation}
 L_{\textbf{K}_{1} \textbf{K}_{2}}(\omega)=L^{0}_{\textbf{K}_{1} \textbf{K}_{2}}(\omega)+L^{0}_{\textbf{K}_{1} \textbf{K}_{3}}(\omega)
 \Xi_{\textbf{K}_{3} \textbf{K}_{4}} L_{\textbf{K}_{4} \textbf{K}_{2}}(\omega), \label{eq9}
\end{equation}
by considering the temperature dependent non-interacting electron-hole Green's function\cite{marini2008}
\begin{equation}
 L^{0}_{\textbf{K}_{1} \textbf{K}_{2}}(\omega,T)=\Big[\frac{f_{c_{1} \textbf{k}_{1}}-f_{v_{1} \textbf{k}_{1}}}{\omega-E_{c_{1} \textbf{k}_{1}}(T)
-E_{v_{1} \textbf{k}_{1}}(T)+0^{+}} \Big]\delta_{\textbf{K}_{1} \textbf{K}_{2}}. \label{eq9.1}
\end{equation}
Here $\textbf{K}=(c,v,\textbf{k})$, comprises the electronic band index and $\Xi=i(W-V)$ represents the BSE kernel composed by the
difference between the static screened and bare Coulomb potential. We adopt a static BSE kernel to describe excitonic effects following ref. 
\onlinecite{kernelstat1} and ref. \onlinecite{kernelstat2} that have shown its accuracy in predicting optical properties in solids.
By solving eq. (\ref{eq9}) with the temperature dependent propagator, the frozen atom BSE Hamiltonian becomes non-Hermitian due to the
presence of imaginary QP energies $E_{n \textbf{k}}$. This is crucial for computing the optical absorption 
since it provides an intrinsic exciton-phonon linewidth which removes the need of including an artificial broadening.
Note that the electron-electron induced linewidths are disregarded in the present case since, as previously shown,\cite{ee1,ee2} 
their contributions in semiconductors are zero for energy windows 2$E_g$ close to the CBM and VBM.
The BSE is calculated by considering 720 bands and 22 Ry energy cutoff in the screened electron-hole potential. 
The optical absorption is computed with the YAMBO code\cite{Marini20091392} using a fine grid with 24000
random \textbf{k}-points with seven valence and seven conduction bands in the $e$-$h$ kernel. The parameter $0^{+}$ in Eq. (\ref{eq9.1}) is set to 5 meV
for numerical reasons.
%%%%%%%%%%%%%FIG1%%%%%%%%%%%%%%%%%%%%%%%%%%%%%%%%%
\begin{figure}[t]
%\centering
%\includegraphics[width=1.0\columnwidth,height=6cm]{fig1}
\includegraphics[width=0.85\columnwidth]{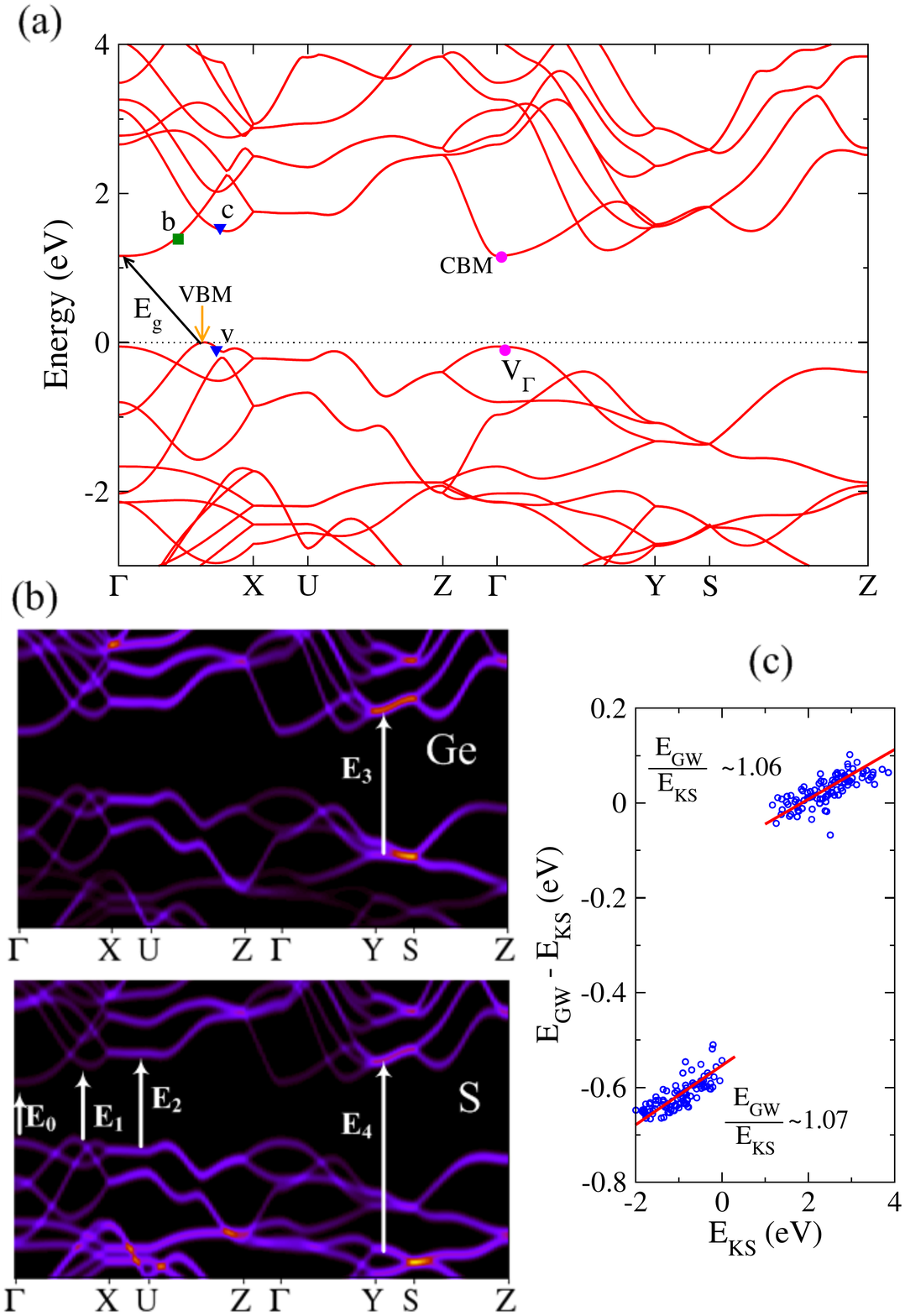}
\caption{\small{(a) Electronic band
structure of GeS showing the indirect band-gap. Blue dots represent the wavevector at which the excitonic state $E_{1}$ takes place
while $V_{\Gamma}$ and $CBM$ are the ones associated to the exciton state $E_{0}$ . (b) $k$-resolved projected
density of states for Ge and S atom contributions. White arrows indicates
the wavevector of the most probable transitions for light polarized in the $x$-direction.
(c) Quasiparticle energy corrections as a function of Khon-Sham energy states showing the renormalization trend. Red lines represent
a linear fit of the data.
}}\label{fig11} 
%{\rule{4cm}{2cm}}
\end{figure}
%%%%%%%%%%%%%FIG1%%%%%%%%%%%%%%%%%%%%%%%%%%%%%%%%%

\section{Results}
We begin our study by describing the electronic structure, shown in Fig. (\ref{fig11}-a) using PBE. There,
we observe an indirect band-gap of $\sim$ 1.16 eV, whose valence band maximum (VBM) is located around three-fifths of the $\Gamma$X path, 
while the conduction band minimum (CBM) is at $\Gamma$. In Fig. (\ref{fig11}-b) we show the $k$-resolved density of states
projected on Ge and S atoms. Overall, the valence (conduction) states are dominated by S (Ge) atoms mostly with $p$-type
contributions. These results are consistent with previous theoretical predictions.\cite{gw-ges,lda-ges}

Note that the direct gap at $\Gamma$ is only 50 meV larger than the indirect gap.
The oscillator strength reveals that the first excitonic state $E_{0}$, is located at $\Gamma$
whereas the wavevector number of the second excitonic state $E_{1}$, occurs from points v to c depicted in Fig (\ref{fig11}-a).
Due to the importance of these optical transitions, our EP coupling study will be conducted at these wavevector numbers.
A schematic representation of the relevant optical transitions in GeS are shown in Fig. (\ref{fig11}-b). 

To properly describe the excitonic effects on the optical properties, one requires a good description of the electronic band-gap.
Thus, we calculate the quasiparticle (QP) corrections within the $G_{0}W_{0}$ approach and plot them as a function of the GGA eigenenergies
in Fig. (\ref{fig11}-c). We obtained a QP correction of 0.58 eV at $\Gamma$ which results in a QP gap of $\approx$ 1.79 eV.
This is consistent with previous GW calculations.\cite{gw-ges}
We also note that the QP corrections are slightly dispersive with respect to the GGA eigenenergies. 
%This prevents to include the GW corrections simply as a rigid shift of the conduction bands. 
In fact, by fitting the QP corrections data to a linear curve, we found that the conduction and valence
bands are, in average, slightly stretched by 6\% and 7\%, respectively. This findings are important since the QP correction at $\Gamma$ and the average
stretching of the bands are taken into account in the form of a scissor operator to formally solve Eq. (\ref{eq9}). 
Notice that 720 bands and an energy cutoff of 18 Ry for the response function are included for evaluating the $G_{0}W_{0}$ corrections.
%%%%%%%%%%%%%FIG1%%%%%%%%%%%%%%%%%%%%%%%%%%%%%%%%%
\begin{figure}%[h]
%\centering
%\includegraphics[width=1.0\columnwidth,height=6cm]{fig1}
\includegraphics[width=1.0\columnwidth]{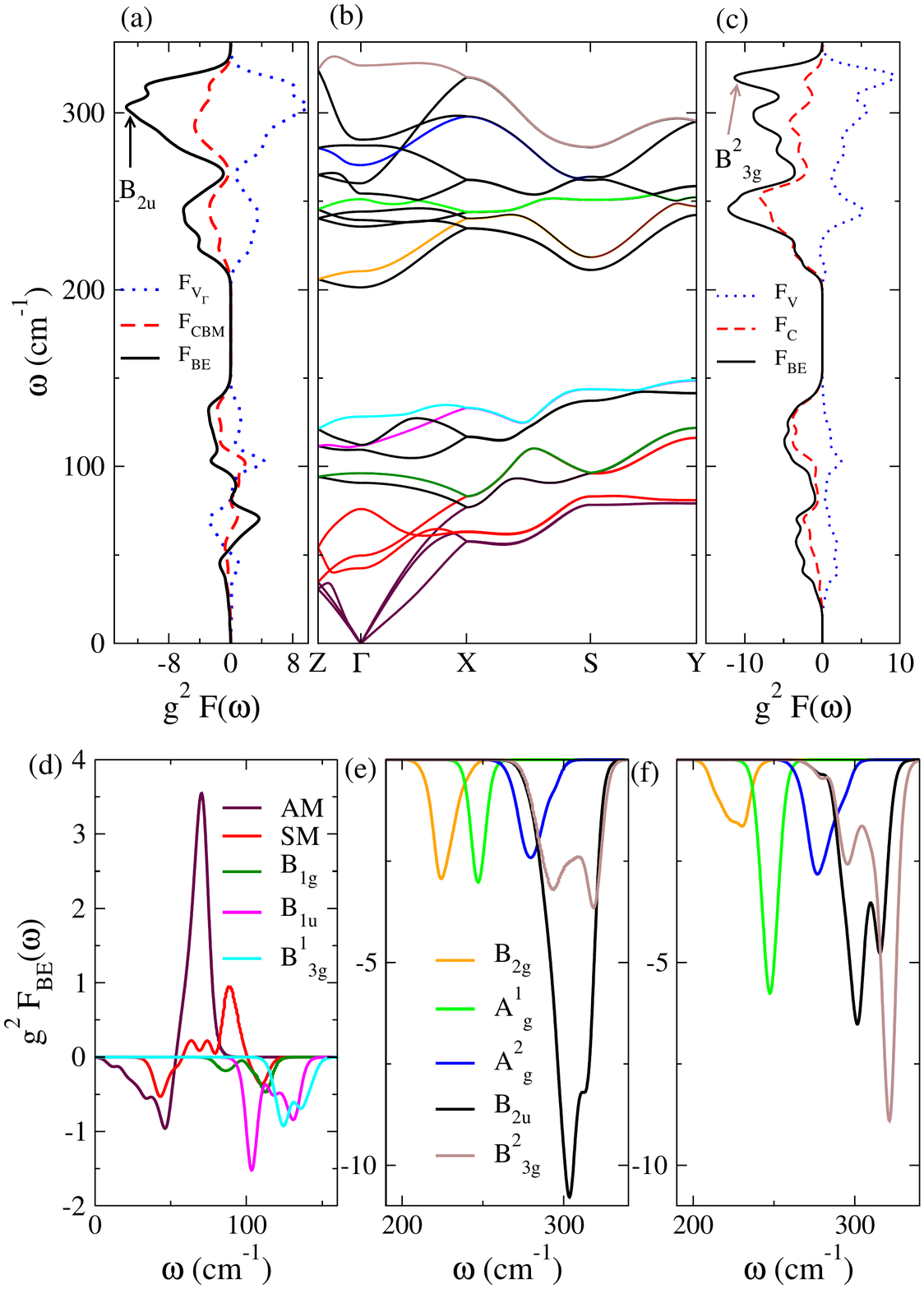}
\caption{\small{(a) Generalized electron-phonon Eliashberg function for $V_{\Gamma}$ (blue dotted-line),
CBM (red dot-dashed line) and band-edge (black solid line). (b) GeS phonon dispersion along selected symmetry points.
(c) Eliashberg function at wavevectors \textbf{c} and \textbf{v} which give rise to the excitonic state $E_1$. The band-edge Eliasberg function,
related to $V_{\Gamma}$ and CBM, projected on
phonon modes with frequencies (d) below 200 cm$^{-1}$ and (e) above 200 cm$^{-1}$. (f) Projected
band-edge Eliashberg function, related to v and c wavevector numbers. Only the most representative 
phonon modes are considered for simplicity. The curves for acoustic and shear modes curves comprise the contribution
of three modes.
%As $\Delta E_{g}(T)\big|_{har}\propto \int d\omega g^{2}F_{n}(\omega)N_{\lambda}(\omega,T)$, the positive $g^{2}F_{n}$ region
%corresponds to a temperature increasing $E_{g}(T)$. As in the anharmonic part the modes of the AR drive the anomaly
%(gray-shadow region).
}}\label{fig2.1} 
%{\rule{4cm}{2cm}}
\end{figure}
%%%%%%%%%%%%%FIG1%%%%%%%%%%%%%%%%%%%%%%%%%%%%%%%%%

We proceed with the study of the EP scattering mechanisms. To this aim, we compute the generalized Eliashberg function 
\begin{equation}
\begin{split}
g^{2}F_{n\textbf{k}}(\omega)=\frac{1}{N}\sum_{n^{'}\textbf{q}\lambda }\Big[\frac{|g^{\textbf{q}\lambda}_{nn'\textbf{k}}|^2}{\epsilon_{n\textbf{k}}
-\epsilon_{n'\textbf{k-q}}}-\frac{1}{2}\frac{\Lambda^{\textbf{q}\lambda}_{nn'\textbf{k}}}
{\epsilon_{n\textbf{k}}-\epsilon_{n'\textbf{k}}}\Big] \\
\times \delta(\omega-\omega_{\textbf{q}\lambda}), \label{eq3}
\end{split}
\end{equation}
which enables us to visualize the EP coupling strength for a given state $|n\textbf{k} \rangle$.
We defined the band-edge Eliashberg function as $F_{BE}$=$F_{c_i}$--$F_{v_i}$, where sub-index
$c_i$($v_i$) refers to a given conduction (valence) state. This function provides useful information regarding
the vibrational modes that eventually couple with an excitonic state rising at the same wavevector numbers.

In Fig. (\ref{fig2.1}-a), we show the Eliashberg function at $\Gamma$, for the highest valence state ($V_\Gamma$) and CBM, the 
wavevectors associated with the exciton state $E_{0}$. At high frequencies $F_{BE}$ present two dominant peaks
spanning from 285 cm$^{-1}$ to 320 cm$^{-1}$ that result mainly from scattering events at $V_\Gamma$.
For frequencies below the phonon dispersion gap, the most important contributions cover the range from 50 to 140 cm$^{-1}$. 
In Fig. (\ref{fig2.1}d-e) we project $F_{BE}$ on each mode. For simplicity, only the most representative modes are shown. Some of 
these modes are schematically represented in Fig. (\ref{fig1}-b).
%%%%%%%%%%%%%FIG1%%%%%%%%%%%%%%%%%%%%%%%%%%%%%%%%%
\begin{figure*}[ht]
%\centering
%\includegraphics[width=1.0\columnwidth,height=6cm]{fig1}
\includegraphics[width=1.35\columnwidth]{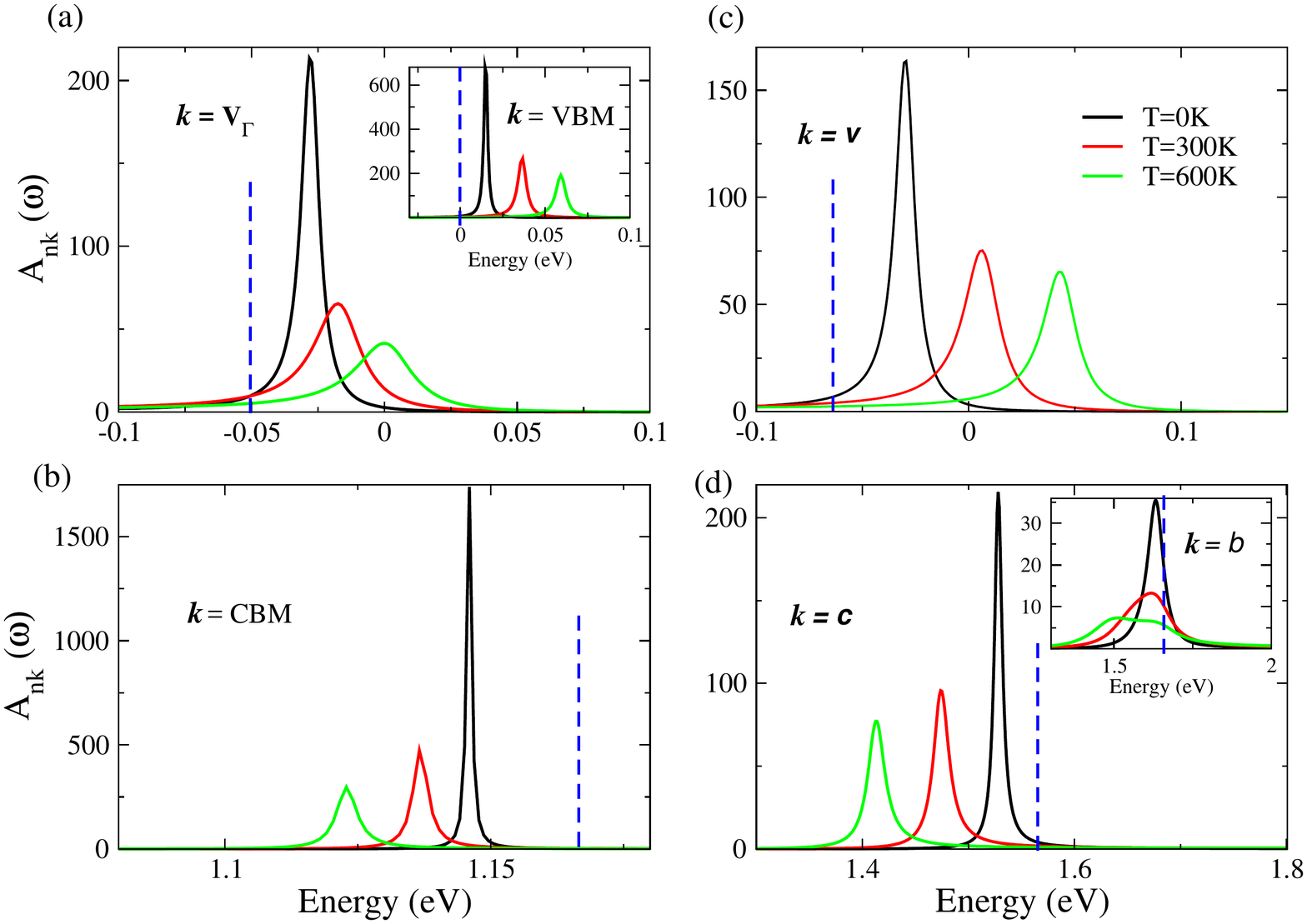}
\caption{\small{Quasiparticle spectral function, $A_{n\textbf{k}}$=$\pi^{-1}$ $\Im{[G_{n\textbf{k}}]}$, for (a) $V_{\Gamma}$ and (b) CBM states. The 
SF for (c) valence state labeled by \textbf{v} and (d) conduction state \textbf{c}, responsible for the excitonic state $E_{1}$.
The insets in (a) and (d) show the SF for wavevectors VBM and $b$ respectively, which are indicated in Fig. (1-c). In particular,
the wavector $k=b$ presents signatures associated to the quasiparticle breakdown.
}}\label{fig3} 
%{\rule{4cm}{2cm}}
\end{figure*}
%%%%%%%%%%%%%FIG1%%%%%%%%%%%%%%%%%%%%%%%%%%%%%%%%%

For frequencies below 90 cm$^{-1}$, the electronic states couple mostly with the acoustic modes (AM) and with less intensity to
the shear modes (SM). Notice that, in contrast to black phosphorus,\cite{villegas1} $F_{BE}$ present non-negligible negative contributions for 
frequencies below 50 cm$^{-1}$, which guarantees the reduction of the band gap as the temperature increases; a feature observed in experiments.\cite{cardonaGeS}
Given that
\begin{equation}
 \Delta E_{g}(T)\propto \int d\omega g^{2}F_{BE}(\omega)[N_{\textbf{q}\lambda}(T)+1/2],
\end{equation}
the Bose-Einstein distribution will always weigh more heavily on the low-frequency region where $F_{BE}$ is negative, causing this way, the monotonic
decreases of the band gap with the temperature.

At higher frequencies, the dominant EP peaks are mostly attributed to
modes B$_{2u}$ and B$^{2}_{3g}$ with B$_{2u}$ providing the most intense contribution, as shown in Fig. (\ref{fig2.1}-e).
Based on this information, one should expect that the first excitonic state $E_{0}$ couples efficiently with the 
infrared longitudinal mode B$_{2u}$. 

Moreover, in Fig. (\ref{fig2.1}-c) we show the EP coupling at ($v$) and ($c$) states, whose wavevector numbers give rise to
the exciton state $E_{1}$. In contrast to the previous case, we note a significant enhancement of the EP strength around 250 cm$^{-1}$
which results from the EP scattering events at $c$ state. In Fig. (\ref{fig2.1}-f) we show the projection of $F_{BE}$ onto
the high frequency optical modes. It clearly shows that the electrons interact more strongly with mode 
B$^{2}_{3g}$. Moreover, the coupling of electrons with mode A$^{1}_{g}$ is considerably enhanced with respect to the excitonic
state $E_{0}$. Therefore, the exciton state $E_{1}$ should couple with the vibrational modes A$^{1}_{g}$, B$^{2}_{3g}$ and B$_{2u}$ which in turn 
will be responsible for the subsequent reduction of the peak intensity and linewidth increase of the optical absorption.
%It should be noticed that the predicted exciton-phonon coupling can be resolved by non-resonant Raman measurements.

From Eq. (\ref{eq6}) in the limit $T\to0$, we estimate the ZPM gap renormalization of $\approx$ $-43\pm3$ meV at $\Gamma$, which
is more than twice the value found in bulk black phosphorus.\cite{villegas1} It should be mentioned that albeit the
similar orthorhombic crystal structure between BP and GeS,
the dominant EP scattering processes at the band-edge of BP are the acoustical, A$^{2}_{g}$ and B$_{2u}$ modes.\cite{villegas1,dreselh}

As previously stated, the MBPT approach allow us to take into account dynamical effects.
Contrary to the single-particle description in which the states hold infinite lifetimes, the QP picture provides finite lifetimes
in the form of the width of a Lorentzian curve centered at $E_{n\textbf{k}}(T)$.

In Fig. (\ref{fig3}a-b) we show the SF related to the $V_{\Gamma}$ and CBM states for different temperatures. The dashed blue lines depict the single
particle (Khon-Sham) energy with an associated infinite lifetime. At $T=0K$, the SF present a Lorentzian shape which is 
particularly sharp at the CBM, reflecting the long QP lifetimes. As the temperature increases, the linewidth gets 
broader and the SF peaks are red-shifted (blue-shifted) for CBM ($V_{\Gamma}$). This shift is a
signature of the shrinking of the energy gap as the temperature increases. For completeness, the inset in Fig. (\ref{fig3}-a) shows the SF at VBM 
which follows similar trends to $V_{\Gamma}$.
%%%%%%%%%%%%%%%%%%%%%%%%%%%%%%%%%%%%%%%%%%%%%%%%%%
\begin{table*}[]
\caption{\small{Representative optical interband transitions energies in GeS at different temperatures for light polarized in the $x$-direction. 
The values are expressed in eV. The experimental values for temperatures below $215K$ are taken from Ref. \onlinecite{cardonaGeS} while for $300K$
from Ref. \onlinecite{cardonaGeS2}}\label{table1}}
 \par
 \begin{center}

 \begin{tabular}{lcccccccccccc}%{0.45\textwidth}[c]
 \hline\hline
 & & &  &  &  &  \\[-5pt]
 \vtop{\hbox{\strut}\hbox{}} & & \multicolumn{2}{c}{$T=0K$} & \multicolumn{2}{c}{$T=84K$} & \multicolumn{2}{c}{$T=215K$} & \multicolumn{2}{c}{$T=300K$} \\[5pt]
  & & Theory & Exp. & Theory  & Exp. & Theory  & Exp.  & Theory  & Exp. \\[0.3ex]
 \hline
 & &  & &  &  & \\
  \ \ \ \ \raisebox{0.3ex}{$E_{0}$} & & \raisebox{0.3ex}{1.702} & \raisebox{0.3ex}{-} & \raisebox{0.3ex}{1.69} &
 \raisebox{0.3ex}{-} & \raisebox{0.3ex}{1.68} & \raisebox{0.3ex}{-} & \raisebox{0.3ex}{1.66} & \raisebox{0.3ex}{1.65\cite{GeSgap}} \\[0.2ex]%
  \ \ \ \ \raisebox{0.3ex}{$E_{1}$} & & \raisebox{0.3ex}{2.143} & \raisebox{0.3ex}{2.144} & \raisebox{0.3ex}{2.128} &
 \raisebox{0.3ex}{2.127} & \raisebox{0.3ex}{2.096} & \raisebox{0.3ex}{2.087} & \raisebox{0.3ex}{2.06} & \raisebox{0.3ex}{2.037} \\[0.2ex]%
%  \ \ \ \ \raisebox{0.3ex}{$E_{3}$} & & \raisebox{0.3ex}{2.81} & \raisebox{0.3ex}{2.832} & \raisebox{0.3ex}{2.815} &
% \raisebox{0.3ex}{2.78} & \raisebox{0.3ex}{2.793} & \raisebox{0.3ex}{2.70} & \raisebox{0.3ex}{2.67} & \raisebox{0.3ex}{2.637} \\[0.2ex]%
  \ \ \ \ \raisebox{0.3ex}{$E_{4}$} & & \raisebox{0.3ex}{3.64} & \raisebox{0.3ex}{3.70} & \raisebox{0.3ex}{3.65} &
 \raisebox{0.3ex}{3.695} & \raisebox{0.3ex}{3.62} & \raisebox{0.3ex}{3.657} & \raisebox{0.3ex}{3.58} & \raisebox{0.3ex}{3.628} \\[0.2ex]%
  \ \ \ \ \raisebox{0.3ex}{$E_{5}$} & & \raisebox{0.3ex}{4.225} & \raisebox{0.3ex}{4.193} & \raisebox{0.3ex}{4.224} &
 \raisebox{0.3ex}{4.162} & \raisebox{0.3ex}{4.18} & \raisebox{0.3ex}{4.089} & \raisebox{0.3ex}{4.09} & \raisebox{0.3ex}{4.031} \\[0.2ex]%

 \hline\hline
 \end{tabular}
 \par
 \end{center}
 \end{table*}
%%%%%%%%%%%%%%%%%%%%%%%%%%%%%%%%%%%%%%%%%%5

%%%%%%%%%%%%FIG1%%%%%%%%%%%%%%%%%%%%%%%%%%%%%%%%%
\begin{figure*}[t]
%\centering
%\includegraphics[width=1.0\columnwidth,height=6cm]{fig1}
\includegraphics[width=1.2\columnwidth]{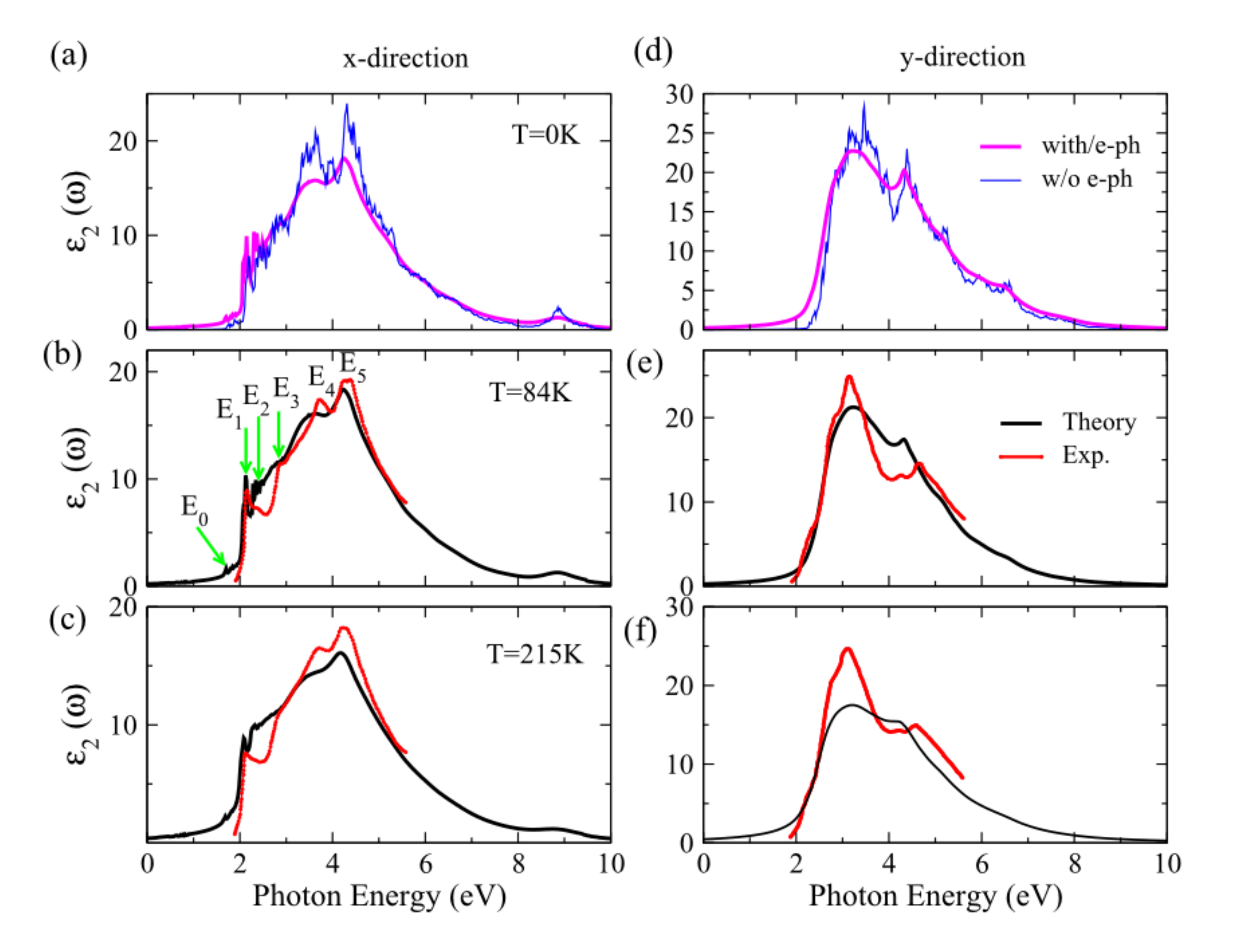}
\caption{\small{Imaginary part of the dielectric function for different light polarizations. Purple and blue lines represent
the theoretical prediction with and without EP corrections, respectively. Red dots are experimental results
for different temperatures extracted from Ref. \onlinecite{cardonaGeS} 
}}\label{fig4} 
%{\rule{4cm}{2cm}}
\end{figure*}
%%%%%%%%%%%%%FIG1%%%%%%%%%%%%%%%%%%%%%%%%%%%%%%%%%

Analogously, in Fig. (\ref{fig3}c-d) we show the SF of the states responsible for the exciton state $E_{1}$.
We observe similar features as in the previous case, regarding the temperature dependence. Note that the SFs
keep their symmetric Lorentzian shape which is a signature of the validity of the QP picture. 
To further discuss this, in the inset of Fig. (\ref{fig3}-d) we show the SF of an
arbitrary point in the conduction band (labeled as $b$ in Fig. (\ref{fig11}-a)).
At $T=0K$, we observe a broad Lorentzian curve that, as the temperature increases, becomes highly asymmetric.
%when is compared with the previous SFs.
For $T=600K$ one can clearly see the formation of a second peak (new state). 
The new state can not be interpreted as a new electron-phonon state. In fact, it has been suggested that the appearance of 
new states are virtual transitions arising
due to energy conservation, and constitute a clear signature of the quasiparticle breakdown \cite{cannu,zpm}.

Finally, we explore the role of the EP interaction on the optical absorption and the relation between our calculations and experimental results. 
In Fig. (\ref{fig4}-a) and Fig. (\ref{fig4}-d) we present the imaginary part of the dielectric function at $T=0K$ for light polarized 
in the $x$ and $y$-direction, respectively. The purple (blue) lines show the spectra with (without) EP contribution. Note that even at $T=0K$, the
region close to the band-edge of the absorption naturally gains a width. Also the position of the peaks including EP corrections
are slightly red-shifted as a consequence of the ZPM renormalization.
In addition, even with the choice of a dense $k$-grid, the absorption without EP contribution shows spiky features over the entire energy 
range. This result clearly highlights the crucial role of including EP interactions
for describing the optical properties of GeS, even at $T=0K$. 

In Fig. (\ref{fig4}b) and  Fig. (\ref{fig4}-c) we show the finite temperature absorption for $x$-polarized light. We clearly observe the natural smoothness 
of the theoretical curves that describe nicely the position of the peaks and their corresponding linewidths without using any external parameter.
For instance, experimental results by Logothetidis \textit{et. al},\cite{cardonaGeS} estimated the position of $E_{1}$ peak  at 2.127 eV and 2.087 eV for $T=84K$ and 
$T=215K$, respectively. Our theoretical calculations locate the same peak
at 2.128 eV and 2.096 eV, in  an excellent agreement. We summarized the energy position of the relevant transitions 
in Table \ref{table1}.
Notice that as the temperature increases the linewidths of our results deviate from the experimental ones, especially for 
the case of light polarized in $y$-direction. We argue that the reason for such deviation might be related to the accurate description of
electronic states entering in Eq. (\ref{eq9.1}). Indeed, by inspecting Eq. (\ref{eq4}) and Eq. (\ref{eq5}), we realized that the
energy dispersion curvature can affect considerably the results, especially the broadening. Thus, a considerable improvement of 
the EP lifetimes could be achieved by computing the EP coupling on top of G$_{0}$W$_{0}$ electronic structure.\cite{gonze} However this procedure
is at this point computationally demanding and thus, not performing in this work. Although this drawbacks, overall, our GGA results
predicts fairly well the optical properties of GeS, especially for $x$-polarized light.
Given the resemblance of the selection rules, band structures and optical absorption curves\cite{GeSgap,GeSoptic} among
the group-VI monochalcogenides, one can expect that, indeed, the EP coupling could play an important role for the description
of the optical properties of this entire family.

Finally, we would like to mention that in few-layers and single-layer GeS forms, one would also expect a monotonic reduction of the band 
gap with the temperature. This can be ascribed by the resemblance of our phonon dispersion results for lower frequencies - the ones responsible
for the gap temperature dependence slope - to recent vibrational studies in monolayer GeS.\cite{GeSmono} Moreover, due to the reduced screening in the few-layer
forms, the EP and exciton-phonon coupling might be enhanced, especially for optical modes.

\section{Conclusions}
We provided a theoretical description of the electron-phonon coupling effects on the electronic and optical
properties of orthorhombic GeS based on \emph{ab-initio} many body perturbation theory. Our results shown 
that, at the band-edge, the longitudinal mode $B_{2u}$ couples efficiently
with the  electronic states. In addition, the electronic states that give rise to the exciton $E_{1}$ couple mostly
with the vibrational modes $A^{1}_{g}$ and $B^{2}_{3g}$. Our results for the optical absorption shown
that even at $T\to0K$, the role of EP coupling is crucial to properly describe the absorption linewithds and peak positions. 
Our findings suggest that in order to properly describe group-VI
monochalcogenides one should include the electron-phonon interaction effects.

\section{Acknowledgements}
CEPV acknowledges A. Molina-S\'{a}nchez for fruitfull discussions and the financial support from the Brazilian agency
FAPESP grant number 2015/14899-0 and 2012/24227-1. AM acknowledges the funding received from 
the {\em Futuro in Ricerca} grant No. RBFR12SW0J of the Italian Ministry of Education, University and Research MIUR,
the European Union project MaX {\em Materials design at the eXascale} H2020-EINFRA-2015-1, Grant agreement n. 676598 and
{\em Nanoscience Foundries and Fine Analysis - Europe} H2020-INFRAIA-2014-2015, Grant agreement n. 654360. A. R. R. acknowledges 
support from ICTP-SAIRF (FAPESP project 2011/11973-4) and the ICTP-Simons Foundation Associate Scheme.
This work uses the computational resources from  GRID-UNESP and CENAPAD/SP.

%\bibliography{paper}

\begin{thebibliography}{61}%
\makeatletter
\providecommand \@ifxundefined [1]{%
 \@ifx{#1\undefined}
}%
\providecommand \@ifnum [1]{%
 \ifnum #1\expandafter \@firstoftwo
 \else \expandafter \@secondoftwo
 \fi
}%
\providecommand \@ifx [1]{%
 \ifx #1\expandafter \@firstoftwo
 \else \expandafter \@secondoftwo
 \fi
}%
\providecommand \natexlab [1]{#1}%
\providecommand \enquote  [1]{``#1''}%
\providecommand \bibnamefont  [1]{#1}%
\providecommand \bibfnamefont [1]{#1}%
\providecommand \citenamefont [1]{#1}%
\providecommand \href@noop [0]{\@secondoftwo}%
\providecommand \href [0]{\begingroup \@sanitize@url \@href}%
\providecommand \@href[1]{\@@startlink{#1}\@@href}%
\providecommand \@@href[1]{\endgroup#1\@@endlink}%
\providecommand \@sanitize@url [0]{\catcode `\\12\catcode `\$12\catcode
  `\&12\catcode `\#12\catcode `\^12\catcode `\_12\catcode `\%12\relax}%
\providecommand \@@startlink[1]{}%
\providecommand \@@endlink[0]{}%
\providecommand \url  [0]{\begingroup\@sanitize@url \@url }%
\providecommand \@url [1]{\endgroup\@href {#1}{\urlprefix }}%
\providecommand \urlprefix  [0]{URL }%
\providecommand \Eprint [0]{\href }%
\providecommand \doibase [0]{http://dx.doi.org/}%
\providecommand \selectlanguage [0]{\@gobble}%
\providecommand \bibinfo  [0]{\@secondoftwo}%
\providecommand \bibfield  [0]{\@secondoftwo}%
\providecommand \translation [1]{[#1]}%
\providecommand \BibitemOpen [0]{}%
\providecommand \bibitemStop [0]{}%
\providecommand \bibitemNoStop [0]{.\EOS\space}%
\providecommand \EOS [0]{\spacefactor3000\relax}%
\providecommand \BibitemShut  [1]{\csname bibitem#1\endcsname}%
\let\auto@bib@innerbib\@empty
%</preamble>
\bibitem [{\citenamefont {Geim}\ and\ \citenamefont
  {Grigorieva}(2013)}]{layer1}%
  \BibitemOpen
  \bibfield  {author} {\bibinfo {author} {\bibfnamefont {A.~K.}\ \bibnamefont
  {Geim}}\ and\ \bibinfo {author} {\bibfnamefont {I.~V.}\ \bibnamefont
  {Grigorieva}},\ }\href {\doibase 10.1038/nature12385} {\bibfield  {journal}
  {\bibinfo  {journal} {Nature}\ }\textbf {\bibinfo {volume} {499}},\ \bibinfo
  {pages} {419} (\bibinfo {year} {2013})}\BibitemShut {NoStop}%
\bibitem [{\citenamefont {Wang}\ \emph {et~al.}(2014)\citenamefont {Wang},
  \citenamefont {Liu}, \citenamefont {Fu}, \citenamefont {Fang}, \citenamefont
  {Zhoue},\ and\ \citenamefont {Liu}}]{layer2}%
  \BibitemOpen
  \bibfield  {author} {\bibinfo {author} {\bibfnamefont {H.}~\bibnamefont
  {Wang}}, \bibinfo {author} {\bibfnamefont {F.}~\bibnamefont {Liu}}, \bibinfo
  {author} {\bibfnamefont {W.}~\bibnamefont {Fu}}, \bibinfo {author}
  {\bibfnamefont {Z.}~\bibnamefont {Fang}}, \bibinfo {author} {\bibfnamefont
  {W.}~\bibnamefont {Zhoue}}, \ and\ \bibinfo {author} {\bibfnamefont
  {Z.}~\bibnamefont {Liu}},\ }\href {\doibase 10.1039/C4NR03435J} {\bibfield
  {journal} {\bibinfo  {journal} {Nanoscale}\ }\textbf {\bibinfo {volume}
  {6}},\ \bibinfo {pages} {12250} (\bibinfo {year} {2014})}\BibitemShut
  {NoStop}%
\bibitem [{\citenamefont {Buscema}\ \emph {et~al.}(2015)\citenamefont
  {Buscema}, \citenamefont {Island}, \citenamefont {Groenendijk}, \citenamefont
  {Blanter}, \citenamefont {Steele}, \citenamefont {van~der Zanta},\ and\
  \citenamefont {Castellanos-Gomez}}]{opto}%
  \BibitemOpen
  \bibfield  {author} {\bibinfo {author} {\bibfnamefont {M.}~\bibnamefont
  {Buscema}}, \bibinfo {author} {\bibfnamefont {J.~O.}\ \bibnamefont {Island}},
  \bibinfo {author} {\bibfnamefont {D.~J.}\ \bibnamefont {Groenendijk}},
  \bibinfo {author} {\bibfnamefont {S.~I.}\ \bibnamefont {Blanter}}, \bibinfo
  {author} {\bibfnamefont {G.~A.}\ \bibnamefont {Steele}}, \bibinfo {author}
  {\bibfnamefont {H.~S.~J.}\ \bibnamefont {van~der Zanta}}, \ and\ \bibinfo
  {author} {\bibfnamefont {A.}~\bibnamefont {Castellanos-Gomez}},\ }\href
  {\doibase 10.1039/C5CS00106D} {\bibfield  {journal} {\bibinfo  {journal}
  {Chem. Soc. Rev.}\ }\textbf {\bibinfo {volume} {44}},\ \bibinfo {pages}
  {3691} (\bibinfo {year} {2015})}\BibitemShut {NoStop}%
\bibitem [{\citenamefont {Furchi}\ \emph {et~al.}(2014)\citenamefont {Furchi},
  \citenamefont {Pospischil}, \citenamefont {Libisch}, \citenamefont
  {Burgd{\"o}rfer},\ and\ \citenamefont {Mueller}}]{photo}%
  \BibitemOpen
  \bibfield  {author} {\bibinfo {author} {\bibfnamefont {M.~M.}\ \bibnamefont
  {Furchi}}, \bibinfo {author} {\bibfnamefont {A.}~\bibnamefont {Pospischil}},
  \bibinfo {author} {\bibfnamefont {F.}~\bibnamefont {Libisch}}, \bibinfo
  {author} {\bibfnamefont {J.}~\bibnamefont {Burgd{\"o}rfer}}, \ and\ \bibinfo
  {author} {\bibfnamefont {T.}~\bibnamefont {Mueller}},\ }\href {\doibase
  10.1021/nl501962c} {\bibfield  {journal} {\bibinfo  {journal} {Nano Lett.}\
  }\textbf {\bibinfo {volume} {14}},\ \bibinfo {pages} {4785–4791} (\bibinfo
  {year} {2014})}\BibitemShut {NoStop}%
\bibitem [{\citenamefont {Britnell}\ \emph {et~al.}(2013)\citenamefont
  {Britnell}, \citenamefont {Ribeiro}, \citenamefont {Eckmann}, \citenamefont
  {Jalil}, \citenamefont {Belle}, \citenamefont {Mishchenko}, \citenamefont
  {Kim}, \citenamefont {Gorbachev}, \citenamefont {Georgiou}, \citenamefont
  {Morozov}, \citenamefont {Grigorenko}, \citenamefont {Geim}, \citenamefont
  {Casiraghi}, \citenamefont {Neto},\ and\ \citenamefont {Novoselov}}]{photo2}%
  \BibitemOpen
  \bibfield  {author} {\bibinfo {author} {\bibfnamefont {L.}~\bibnamefont
  {Britnell}}, \bibinfo {author} {\bibfnamefont {R.~M.}\ \bibnamefont
  {Ribeiro}}, \bibinfo {author} {\bibfnamefont {A.}~\bibnamefont {Eckmann}},
  \bibinfo {author} {\bibfnamefont {R.}~\bibnamefont {Jalil}}, \bibinfo
  {author} {\bibfnamefont {B.~D.}\ \bibnamefont {Belle}}, \bibinfo {author}
  {\bibfnamefont {A.}~\bibnamefont {Mishchenko}}, \bibinfo {author}
  {\bibfnamefont {Y.-J.}\ \bibnamefont {Kim}}, \bibinfo {author} {\bibfnamefont
  {R.~V.}\ \bibnamefont {Gorbachev}}, \bibinfo {author} {\bibfnamefont
  {T.}~\bibnamefont {Georgiou}}, \bibinfo {author} {\bibfnamefont {S.~V.}\
  \bibnamefont {Morozov}}, \bibinfo {author} {\bibfnamefont {A.~N.}\
  \bibnamefont {Grigorenko}}, \bibinfo {author} {\bibfnamefont {A.~K.}\
  \bibnamefont {Geim}}, \bibinfo {author} {\bibfnamefont {C.}~\bibnamefont
  {Casiraghi}}, \bibinfo {author} {\bibfnamefont {A.~H.~C.}\ \bibnamefont
  {Neto}}, \ and\ \bibinfo {author} {\bibfnamefont {K.~S.}\ \bibnamefont
  {Novoselov}},\ }\href {\doibase 10.1126/science.1235547} {\bibfield
  {journal} {\bibinfo  {journal} {Science}\ }\textbf {\bibinfo {volume}
  {340}},\ \bibinfo {pages} {1311–1314} (\bibinfo {year} {2013})}\BibitemShut
  {NoStop}%
\bibitem [{\citenamefont {Zhao}\ \emph {et~al.}(2014)\citenamefont {Zhao},
  \citenamefont {Lo}, \citenamefont {Zhang}, \citenamefont {Sun}, \citenamefont
  {Tan}, \citenamefont {Uher}, \citenamefont {Dravid},\ and\ \citenamefont
  {Kanatzidis}}]{thermo}%
  \BibitemOpen
  \bibfield  {author} {\bibinfo {author} {\bibfnamefont {L.-D.}\ \bibnamefont
  {Zhao}}, \bibinfo {author} {\bibfnamefont {S.-H.}\ \bibnamefont {Lo}},
  \bibinfo {author} {\bibfnamefont {Y.}~\bibnamefont {Zhang}}, \bibinfo
  {author} {\bibfnamefont {H.}~\bibnamefont {Sun}}, \bibinfo {author}
  {\bibfnamefont {G.}~\bibnamefont {Tan}}, \bibinfo {author} {\bibfnamefont
  {C.}~\bibnamefont {Uher}}, \bibinfo {author} {\bibfnamefont {C.~W. V.~P.}\
  \bibnamefont {Dravid}}, \ and\ \bibinfo {author} {\bibfnamefont {M.~G.}\
  \bibnamefont {Kanatzidis}},\ }\href {\doibase 10.1038/nature13184} {\bibfield
   {journal} {\bibinfo  {journal} {Nature}\ }\textbf {\bibinfo {volume}
  {508}},\ \bibinfo {pages} {373–377} (\bibinfo {year} {2014})}\BibitemShut
  {NoStop}%
\bibitem [{\citenamefont {Ding}\ \emph {et~al.}(2015)\citenamefont {Ding},
  \citenamefont {Gao},\ and\ \citenamefont {Yao}}]{thermo2}%
  \BibitemOpen
  \bibfield  {author} {\bibinfo {author} {\bibfnamefont {G.}~\bibnamefont
  {Ding}}, \bibinfo {author} {\bibfnamefont {G.}~\bibnamefont {Gao}}, \ and\
  \bibinfo {author} {\bibfnamefont {K.}~\bibnamefont {Yao}},\ }\href {\doibase
  10.1038/srep09567} {\bibfield  {journal} {\bibinfo  {journal} {Sci. Rep.}\
  }\textbf {\bibinfo {volume} {5}},\ \bibinfo {pages} {9567} (\bibinfo {year}
  {2015})}\BibitemShut {NoStop}%
\bibitem [{\citenamefont {Senguptaa}\ \emph {et~al.}(2011)\citenamefont
  {Senguptaa}, \citenamefont {Bhattacharyaa}, \citenamefont {Bandyopadhyayb},\
  and\ \citenamefont {Bhowmick}}]{graph}%
  \BibitemOpen
  \bibfield  {author} {\bibinfo {author} {\bibfnamefont {R.}~\bibnamefont
  {Senguptaa}}, \bibinfo {author} {\bibfnamefont {M.}~\bibnamefont
  {Bhattacharyaa}}, \bibinfo {author} {\bibfnamefont {S.}~\bibnamefont
  {Bandyopadhyayb}}, \ and\ \bibinfo {author} {\bibfnamefont {A.~K.}\
  \bibnamefont {Bhowmick}},\ }\href {\doibase 10.1021/nn500064s} {\bibfield
  {journal} {\bibinfo  {journal} {Prog. Polym. Sci.}\ }\textbf {\bibinfo
  {volume} {36}},\ \bibinfo {pages} {638–670} (\bibinfo {year}
  {2011})}\BibitemShut {NoStop}%
\bibitem [{\citenamefont {Paine}\ and\ \citenamefont {Narula}(1990)}]{bn1}%
  \BibitemOpen
  \bibfield  {author} {\bibinfo {author} {\bibfnamefont {R.~T.}\ \bibnamefont
  {Paine}}\ and\ \bibinfo {author} {\bibfnamefont {C.~K.}\ \bibnamefont
  {Narula}},\ }\href {\doibase 10.1021/cr00099a004} {\bibfield  {journal}
  {\bibinfo  {journal} {Chem. Rev.}\ }\textbf {\bibinfo {volume} {90}},\
  \bibinfo {pages} {73–91} (\bibinfo {year} {1990})}\BibitemShut {NoStop}%
\bibitem [{\citenamefont {Pakdel}\ \emph {et~al.}(2014)\citenamefont {Pakdel},
  \citenamefont {Bandoa},\ and\ \citenamefont {Golberg}}]{bn2}%
  \BibitemOpen
  \bibfield  {author} {\bibinfo {author} {\bibfnamefont {A.}~\bibnamefont
  {Pakdel}}, \bibinfo {author} {\bibfnamefont {Y.}~\bibnamefont {Bandoa}}, \
  and\ \bibinfo {author} {\bibfnamefont {D.}~\bibnamefont {Golberg}},\ }\href
  {\doibase 10.1039/C3CS60260E} {\bibfield  {journal} {\bibinfo  {journal}
  {Chem. Soc. Rev.}\ }\textbf {\bibinfo {volume} {43}},\ \bibinfo {pages} {934}
  (\bibinfo {year} {2014})}\BibitemShut {NoStop}%
\bibitem [{\citenamefont {Jariwala}\ \emph {et~al.}(2014)\citenamefont
  {Jariwala}, \citenamefont {Sangwan}, \citenamefont {Lauhon}, \citenamefont
  {Marks},\ and\ \citenamefont {Hersam}}]{tmdc1}%
  \BibitemOpen
  \bibfield  {author} {\bibinfo {author} {\bibfnamefont {D.}~\bibnamefont
  {Jariwala}}, \bibinfo {author} {\bibfnamefont {V.~K.}\ \bibnamefont
  {Sangwan}}, \bibinfo {author} {\bibfnamefont {L.~J.}\ \bibnamefont {Lauhon}},
  \bibinfo {author} {\bibfnamefont {T.~J.}\ \bibnamefont {Marks}}, \ and\
  \bibinfo {author} {\bibfnamefont {M.~C.}\ \bibnamefont {Hersam}},\ }\href
  {\doibase 10.1021/nn500064s} {\bibfield  {journal} {\bibinfo  {journal} {ACS
  Nano}\ }\textbf {\bibinfo {volume} {8}},\ \bibinfo {pages} {1102–1120}
  (\bibinfo {year} {2014})}\BibitemShut {NoStop}%
\bibitem [{\citenamefont {Wang}\ \emph {et~al.}(2012)\citenamefont {Wang},
  \citenamefont {Kalantar-Zadeh}, \citenamefont {Kis}, \citenamefont
  {Coleman},\ and\ \citenamefont {Strano}}]{tmdc2}%
  \BibitemOpen
  \bibfield  {author} {\bibinfo {author} {\bibfnamefont {Q.~H.}\ \bibnamefont
  {Wang}}, \bibinfo {author} {\bibfnamefont {K.}~\bibnamefont
  {Kalantar-Zadeh}}, \bibinfo {author} {\bibfnamefont {A.}~\bibnamefont {Kis}},
  \bibinfo {author} {\bibfnamefont {J.~N.}\ \bibnamefont {Coleman}}, \ and\
  \bibinfo {author} {\bibfnamefont {M.~S.}\ \bibnamefont {Strano}},\ }\href
  {\doibase 10.1038/nnano.2012.193} {\bibfield  {journal} {\bibinfo  {journal}
  {Nat. Nanotech.}\ }\textbf {\bibinfo {volume} {7}},\ \bibinfo {pages}
  {699–712} (\bibinfo {year} {2012})}\BibitemShut {NoStop}%
\bibitem [{\citenamefont {Castellanos-Gomez}(2015)}]{BPrev}%
  \BibitemOpen
  \bibfield  {author} {\bibinfo {author} {\bibfnamefont {A.}~\bibnamefont
  {Castellanos-Gomez}},\ }\href {\doibase 10.1021/acs.jpclett.5b01686}
  {\bibfield  {journal} {\bibinfo  {journal} {Jour. Phys. Chem. Lett.}\
  }\textbf {\bibinfo {volume} {6}},\ \bibinfo {pages} {4280} (\bibinfo {year}
  {2015})}\BibitemShut {NoStop}%
\bibitem [{\citenamefont {Ling}\ \emph {et~al.}(2015)\citenamefont {Ling},
  \citenamefont {Wang}, \citenamefont {Huang}, \citenamefont {Xia},\ and\
  \citenamefont {Dresselhaus}}]{BPrev2}%
  \BibitemOpen
  \bibfield  {author} {\bibinfo {author} {\bibfnamefont {X.}~\bibnamefont
  {Ling}}, \bibinfo {author} {\bibfnamefont {H.}~\bibnamefont {Wang}}, \bibinfo
  {author} {\bibfnamefont {S.}~\bibnamefont {Huang}}, \bibinfo {author}
  {\bibfnamefont {F.}~\bibnamefont {Xia}}, \ and\ \bibinfo {author}
  {\bibfnamefont {M.~S.}\ \bibnamefont {Dresselhaus}},\ }\href@noop {}
  {\bibfield  {journal} {\bibinfo  {journal} {Proc. Natl. Acad. Sci}\ }\textbf
  {\bibinfo {volume} {112}},\ \bibinfo {pages} {201416581} (\bibinfo {year}
  {2015})}\BibitemShut {NoStop}%
\bibitem [{\citenamefont {Gomes}\ and\ \citenamefont {Carvalho}(2015)}]{mono1}%
  \BibitemOpen
  \bibfield  {author} {\bibinfo {author} {\bibfnamefont {L.~C.}\ \bibnamefont
  {Gomes}}\ and\ \bibinfo {author} {\bibfnamefont {A.}~\bibnamefont
  {Carvalho}},\ }\href {\doibase 10.1103/PhysRevB.92.085406} {\bibfield
  {journal} {\bibinfo  {journal} {Phys. Rev. B}\ }\textbf {\bibinfo {volume}
  {92}},\ \bibinfo {pages} {085406} (\bibinfo {year} {2015})}\BibitemShut
  {NoStop}%
\bibitem [{\citenamefont {Singh}\ and\ \citenamefont {Hennig}(2014)}]{mono2}%
  \BibitemOpen
  \bibfield  {author} {\bibinfo {author} {\bibfnamefont {A.~K.}\ \bibnamefont
  {Singh}}\ and\ \bibinfo {author} {\bibfnamefont {R.~G.}\ \bibnamefont
  {Hennig}},\ }\href {\doibase 10.1063/1.4891230} {\bibfield  {journal}
  {\bibinfo  {journal} {Appl. Phys. Lett.}\ }\textbf {\bibinfo {volume}
  {105}},\ \bibinfo {pages} {042103} (\bibinfo {year} {2014})}\BibitemShut
  {NoStop}%
\bibitem [{\citenamefont {Ribeiro}\ \emph {et~al.}(2016)\citenamefont
  {Ribeiro}, \citenamefont {Villegas}, \citenamefont {Bahamon}, \citenamefont
  {Muraca}, \citenamefont {Castro-Neto}, \citenamefont {de~Souza},
  \citenamefont {Rocha}, \citenamefont {Pimenta},\ and\ \citenamefont
  {de~Matos}}]{villegas2}%
  \BibitemOpen
  \bibfield  {author} {\bibinfo {author} {\bibfnamefont {H.~B.}\ \bibnamefont
  {Ribeiro}}, \bibinfo {author} {\bibfnamefont {C.~E.~P.}\ \bibnamefont
  {Villegas}}, \bibinfo {author} {\bibfnamefont {D.~A.}\ \bibnamefont
  {Bahamon}}, \bibinfo {author} {\bibfnamefont {D.}~\bibnamefont {Muraca}},
  \bibinfo {author} {\bibfnamefont {A.~H.}\ \bibnamefont {Castro-Neto}},
  \bibinfo {author} {\bibfnamefont {E.~A.~T.}\ \bibnamefont {de~Souza}},
  \bibinfo {author} {\bibfnamefont {A.~R.}\ \bibnamefont {Rocha}}, \bibinfo
  {author} {\bibfnamefont {M.~A.}\ \bibnamefont {Pimenta}}, \ and\ \bibinfo
  {author} {\bibfnamefont {C.~J.~S.}\ \bibnamefont {de~Matos}},\ }\href
  {\doibase 10.1038/ncomms12191} {\bibfield  {journal} {\bibinfo  {journal}
  {Nat. Commun.}\ }\textbf {\bibinfo {volume} {7}},\ \bibinfo {pages} {12191}
  (\bibinfo {year} {2016})}\BibitemShut {NoStop}%
\bibitem [{\citenamefont {Villegas}\ \emph
  {et~al.}(2016{\natexlab{a}})\citenamefont {Villegas}, \citenamefont {Rodin},
  \citenamefont {Carvalho},\ and\ \citenamefont {Rocha}}]{villegasexci}%
  \BibitemOpen
  \bibfield  {author} {\bibinfo {author} {\bibfnamefont {C.~E.~P.}\
  \bibnamefont {Villegas}}, \bibinfo {author} {\bibfnamefont {A.}~\bibnamefont
  {Rodin}}, \bibinfo {author} {\bibfnamefont {A.}~\bibnamefont {Carvalho}}, \
  and\ \bibinfo {author} {\bibfnamefont {A.~R.}\ \bibnamefont {Rocha}},\ }\href
  {\doibase 10.1039/C6CP05566D} {\bibfield  {journal} {\bibinfo  {journal}
  {Phys. Chem. Chem. Phys.}\ }\textbf {\bibinfo {volume} {18}},\ \bibinfo
  {pages} {27829} (\bibinfo {year} {2016}{\natexlab{a}})}\BibitemShut {NoStop}%
\bibitem [{\citenamefont {Xia}\ \emph {et~al.}(2015)\citenamefont {Xia},
  \citenamefont {Wang},\ and\ \citenamefont {Jia}}]{aniso1}%
  \BibitemOpen
  \bibfield  {author} {\bibinfo {author} {\bibfnamefont {F.}~\bibnamefont
  {Xia}}, \bibinfo {author} {\bibfnamefont {H.}~\bibnamefont {Wang}}, \ and\
  \bibinfo {author} {\bibfnamefont {Y.}~\bibnamefont {Jia}},\ }\href {\doibase
  10.1038/ncomms5458} {\bibfield  {journal} {\bibinfo  {journal} {Nat.
  Commun.}\ }\textbf {\bibinfo {volume} {5}},\ \bibinfo {pages} {4458}
  (\bibinfo {year} {2015})}\BibitemShut {NoStop}%
\bibitem [{\citenamefont {Qin}\ \emph {et~al.}(2016)\citenamefont {Qin},
  \citenamefont {Qin}, \citenamefont {Fang}, \citenamefont {Zhang},
  \citenamefont {Yue}, \citenamefont {Yan}, \citenamefont {Hu},\ and\
  \citenamefont {Su}}]{mono-ani}%
  \BibitemOpen
  \bibfield  {author} {\bibinfo {author} {\bibfnamefont {G.}~\bibnamefont
  {Qin}}, \bibinfo {author} {\bibfnamefont {Z.}~\bibnamefont {Qin}}, \bibinfo
  {author} {\bibfnamefont {W.-Z.}\ \bibnamefont {Fang}}, \bibinfo {author}
  {\bibfnamefont {L.-C.}\ \bibnamefont {Zhang}}, \bibinfo {author}
  {\bibfnamefont {S.-Y.}\ \bibnamefont {Yue}}, \bibinfo {author} {\bibfnamefont
  {Q.-B.}\ \bibnamefont {Yan}}, \bibinfo {author} {\bibfnamefont
  {M.}~\bibnamefont {Hu}}, \ and\ \bibinfo {author} {\bibfnamefont
  {G.}~\bibnamefont {Su}},\ }\href {\doibase 10.1039/C6NR01349J} {\bibfield
  {journal} {\bibinfo  {journal} {Nanoscale}\ }\textbf {\bibinfo {volume}
  {8}},\ \bibinfo {pages} {11306} (\bibinfo {year} {2016})}\BibitemShut
  {NoStop}%
\bibitem [{\citenamefont {Villegas}\ \emph
  {et~al.}(2016{\natexlab{b}})\citenamefont {Villegas}, \citenamefont {Rocha},\
  and\ \citenamefont {Marini}}]{villegas1}%
  \BibitemOpen
  \bibfield  {author} {\bibinfo {author} {\bibfnamefont {C.~E.~P.}\
  \bibnamefont {Villegas}}, \bibinfo {author} {\bibfnamefont {A.~R.}\
  \bibnamefont {Rocha}}, \ and\ \bibinfo {author} {\bibfnamefont
  {A.}~\bibnamefont {Marini}},\ }\href {\doibase 10.1021/acs.nanolett.6b02035}
  {\bibfield  {journal} {\bibinfo  {journal} {Nano Lett.}\ }\textbf {\bibinfo
  {volume} {16}},\ \bibinfo {pages} {5095} (\bibinfo {year}
  {2016}{\natexlab{b}})}\BibitemShut {NoStop}%
\bibitem [{\citenamefont {Eymard}\ and\ \citenamefont {Otto}(1977)}]{GeSgap}%
  \BibitemOpen
  \bibfield  {author} {\bibinfo {author} {\bibfnamefont {R.}~\bibnamefont
  {Eymard}}\ and\ \bibinfo {author} {\bibfnamefont {A.}~\bibnamefont {Otto}},\
  }\href {\doibase 10.1103/PhysRevB.16.1616} {\bibfield  {journal} {\bibinfo
  {journal} {Phys. Rev. B}\ }\textbf {\bibinfo {volume} {16}},\ \bibinfo
  {pages} {1616} (\bibinfo {year} {1977})}\BibitemShut {NoStop}%
\bibitem [{\citenamefont {Latiff}\ \emph {et~al.}(2015)\citenamefont {Latiff},
  \citenamefont {Teo}, \citenamefont {Sofer}, \citenamefont {Huber},
  \citenamefont {Fisherc},\ and\ \citenamefont {Pumera}}]{toxi}%
  \BibitemOpen
  \bibfield  {author} {\bibinfo {author} {\bibfnamefont {N.}~\bibnamefont
  {Latiff}}, \bibinfo {author} {\bibfnamefont {W.~Z.}\ \bibnamefont {Teo}},
  \bibinfo {author} {\bibfnamefont {Z.}~\bibnamefont {Sofer}}, \bibinfo
  {author} {\bibfnamefont {S.}~\bibnamefont {Huber}}, \bibinfo {author}
  {\bibfnamefont {A.~C.}\ \bibnamefont {Fisherc}}, \ and\ \bibinfo {author}
  {\bibfnamefont {M.}~\bibnamefont {Pumera}},\ }\href {\doibase
  10.1039/C5RA09404F} {\bibfield  {journal} {\bibinfo  {journal} {RSC Adv.}\
  }\textbf {\bibinfo {volume} {5}},\ \bibinfo {pages} {67485} (\bibinfo {year}
  {2015})}\BibitemShut {NoStop}%
\bibitem [{\citenamefont {Lattif}\ \emph {et~al.}(2015)\citenamefont {Lattif},
  \citenamefont {Teo}, \citenamefont {Sofer}, \citenamefont {Fisher},\ and\
  \citenamefont {Pumera}}]{toxibp}%
  \BibitemOpen
  \bibfield  {author} {\bibinfo {author} {\bibfnamefont {N.~M.}\ \bibnamefont
  {Lattif}}, \bibinfo {author} {\bibfnamefont {W.~Z.}\ \bibnamefont {Teo}},
  \bibinfo {author} {\bibfnamefont {Z.}~\bibnamefont {Sofer}}, \bibinfo
  {author} {\bibfnamefont {A.~C.}\ \bibnamefont {Fisher}}, \ and\ \bibinfo
  {author} {\bibfnamefont {M.}~\bibnamefont {Pumera}},\ }\href {\doibase
  10.1002/chem.201502006} {\bibfield  {journal} {\bibinfo  {journal}
  {Chem.-Eur. J.}\ }\textbf {\bibinfo {volume} {21}},\ \bibinfo {pages} {13991}
  (\bibinfo {year} {2015})}\BibitemShut {NoStop}%
\bibitem [{\citenamefont {Favron}\ \emph {et~al.}(2015)\citenamefont {Favron},
  \citenamefont {Gaufr\`es}, \citenamefont {Fossard}, \citenamefont
  {Phaneuf-L'Heureux}, \citenamefont {Tang}, \citenamefont {L\'evesque},
  \citenamefont {Loiseau}, \citenamefont {Leonelli}, \citenamefont
  {Francoeur},\ and\ \citenamefont {Martel}}]{stabp}%
  \BibitemOpen
  \bibfield  {author} {\bibinfo {author} {\bibfnamefont {A.}~\bibnamefont
  {Favron}}, \bibinfo {author} {\bibfnamefont {E.}~\bibnamefont {Gaufr\`es}},
  \bibinfo {author} {\bibfnamefont {F.}~\bibnamefont {Fossard}}, \bibinfo
  {author} {\bibfnamefont {A.-L.}\ \bibnamefont {Phaneuf-L'Heureux}}, \bibinfo
  {author} {\bibfnamefont {N.~Y.-W.}\ \bibnamefont {Tang}}, \bibinfo {author}
  {\bibfnamefont {P.~L.}\ \bibnamefont {L\'evesque}}, \bibinfo {author}
  {\bibfnamefont {A.}~\bibnamefont {Loiseau}}, \bibinfo {author} {\bibfnamefont
  {R.}~\bibnamefont {Leonelli}}, \bibinfo {author} {\bibfnamefont
  {S.}~\bibnamefont {Francoeur}}, \ and\ \bibinfo {author} {\bibfnamefont
  {R.}~\bibnamefont {Martel}},\ }\href {\doibase 10.1038/nmat4299} {\bibfield
  {journal} {\bibinfo  {journal} {Nat. Mat.}\ }\textbf {\bibinfo {volume}
  {14}},\ \bibinfo {pages} {826–832} (\bibinfo {year} {2015})}\BibitemShut
  {NoStop}%
\bibitem [{\citenamefont {Ulaganathan}\ \emph {et~al.}(2016)\citenamefont
  {Ulaganathan}, \citenamefont {Lu}, \citenamefont {Kuo}, \citenamefont
  {Tamalampudi}, \citenamefont {Sankar}, \citenamefont {Boopathi},
  \citenamefont {Anand}, \citenamefont {Yadavm}, \citenamefont {Mathew},
  \citenamefont {Liu}, \citenamefont {Choue},\ and\ \citenamefont
  {Chen}}]{GeSprop1}%
  \BibitemOpen
  \bibfield  {author} {\bibinfo {author} {\bibfnamefont {R.~K.}\ \bibnamefont
  {Ulaganathan}}, \bibinfo {author} {\bibfnamefont {Y.-Y.}\ \bibnamefont {Lu}},
  \bibinfo {author} {\bibfnamefont {C.-J.}\ \bibnamefont {Kuo}}, \bibinfo
  {author} {\bibfnamefont {S.~R.}\ \bibnamefont {Tamalampudi}}, \bibinfo
  {author} {\bibfnamefont {R.}~\bibnamefont {Sankar}}, \bibinfo {author}
  {\bibfnamefont {K.~M.}\ \bibnamefont {Boopathi}}, \bibinfo {author}
  {\bibfnamefont {A.}~\bibnamefont {Anand}}, \bibinfo {author} {\bibfnamefont
  {K.}~\bibnamefont {Yadavm}}, \bibinfo {author} {\bibfnamefont {R.~J.}\
  \bibnamefont {Mathew}}, \bibinfo {author} {\bibfnamefont {C.-R.}\
  \bibnamefont {Liu}}, \bibinfo {author} {\bibfnamefont {F.~C.}\ \bibnamefont
  {Choue}}, \ and\ \bibinfo {author} {\bibfnamefont {Y.-T.}\ \bibnamefont
  {Chen}},\ }\href {\doibase 10.1039/C5NR05988G} {\bibfield  {journal}
  {\bibinfo  {journal} {Nanoscale}\ }\textbf {\bibinfo {volume} {8}},\ \bibinfo
  {pages} {2284} (\bibinfo {year} {2016})}\BibitemShut {NoStop}%
\bibitem [{\citenamefont {Logothetidis}\ \emph {et~al.}(1986)\citenamefont
  {Logothetidis}, \citenamefont {Lautenschlager},\ and\ \citenamefont
  {Cardona}}]{cardonaGeS}%
  \BibitemOpen
  \bibfield  {author} {\bibinfo {author} {\bibfnamefont {S.}~\bibnamefont
  {Logothetidis}}, \bibinfo {author} {\bibfnamefont {P.}~\bibnamefont
  {Lautenschlager}}, \ and\ \bibinfo {author} {\bibfnamefont {M.}~\bibnamefont
  {Cardona}},\ }\href {\doibase 10.1103/PhysRevB.33.1110} {\bibfield  {journal}
  {\bibinfo  {journal} {Phys. Rev. B}\ }\textbf {\bibinfo {volume} {33}},\
  \bibinfo {pages} {1110} (\bibinfo {year} {1986})}\BibitemShut {NoStop}%
\bibitem [{\citenamefont {Hsueh}\ \emph {et~al.}(1996)\citenamefont {Hsueh},
  \citenamefont {Warren}, \citenamefont {Vass}, \citenamefont {Ackland},
  \citenamefont {Clark},\ and\ \citenamefont {Crain}}]{GeSvibra}%
  \BibitemOpen
  \bibfield  {author} {\bibinfo {author} {\bibfnamefont {H.~C.}\ \bibnamefont
  {Hsueh}}, \bibinfo {author} {\bibfnamefont {M.~C.}\ \bibnamefont {Warren}},
  \bibinfo {author} {\bibfnamefont {H.}~\bibnamefont {Vass}}, \bibinfo {author}
  {\bibfnamefont {G.~J.}\ \bibnamefont {Ackland}}, \bibinfo {author}
  {\bibfnamefont {S.~J.}\ \bibnamefont {Clark}}, \ and\ \bibinfo {author}
  {\bibfnamefont {J.}~\bibnamefont {Crain}},\ }\href@noop {} {\bibfield
  {journal} {\bibinfo  {journal} {Phys. rev. B}\ }\textbf {\bibinfo {volume}
  {53}},\ \bibinfo {pages} {14806} (\bibinfo {year} {1996})}\BibitemShut
  {NoStop}%
\bibitem [{\citenamefont {Ch\'ab}\ and\ \citenamefont
  {Bartos}(1984)}]{GeSarpes}%
  \BibitemOpen
  \bibfield  {author} {\bibinfo {author} {\bibfnamefont {V.}~\bibnamefont
  {Ch\'ab}}\ and\ \bibinfo {author} {\bibfnamefont {I.}~\bibnamefont
  {Bartos}},\ }\href {\doibase 10.1002/pssb.2221210132} {\bibfield  {journal}
  {\bibinfo  {journal} {Phys. Stat. Sol. (b)}\ }\textbf {\bibinfo {volume}
  {121}},\ \bibinfo {pages} {301–306} (\bibinfo {year} {1984})}\BibitemShut
  {NoStop}%
\bibitem [{\citenamefont {F.~Lukes}(1981)}]{GeSopt3}%
  \BibitemOpen
  \bibfield  {author} {\bibinfo {author} {\bibfnamefont {A.~L.}\ \bibnamefont
  {F.~Lukes}, \bibfnamefont {E.~Schmidt}},\ }\href {\doibase
  10.1016/0038-1098(81)90038-7} {\bibfield  {journal} {\bibinfo  {journal}
  {Phys. Stat. Sol. (b)}\ }\textbf {\bibinfo {volume} {39}},\ \bibinfo {pages}
  {921} (\bibinfo {year} {1981})}\BibitemShut {NoStop}%
\bibitem [{\citenamefont {Makinistian}\ and\ \citenamefont
  {Albanesi}(2011)}]{GeSoptic}%
  \BibitemOpen
  \bibfield  {author} {\bibinfo {author} {\bibfnamefont {L.}~\bibnamefont
  {Makinistian}}\ and\ \bibinfo {author} {\bibfnamefont {E.~A.}\ \bibnamefont
  {Albanesi}},\ }\href@noop {} {\bibfield  {journal} {\bibinfo  {journal}
  {Comp. Mat. Sci.}\ }\textbf {\bibinfo {volume} {50}},\ \bibinfo {pages}
  {2872} (\bibinfo {year} {2011})}\BibitemShut {NoStop}%
\bibitem [{\citenamefont {Tuttle}\ \emph {et~al.}(2015)\citenamefont {Tuttle},
  \citenamefont {Alhassan},\ and\ \citenamefont {Pantelides}}]{GeSopt4}%
  \BibitemOpen
  \bibfield  {author} {\bibinfo {author} {\bibfnamefont {B.~R.}\ \bibnamefont
  {Tuttle}}, \bibinfo {author} {\bibfnamefont {S.~M.}\ \bibnamefont
  {Alhassan}}, \ and\ \bibinfo {author} {\bibfnamefont {S.~T.}\ \bibnamefont
  {Pantelides}},\ }\href {\doibase 10.1103/PhysRevB.92.235405} {\bibfield
  {journal} {\bibinfo  {journal} {Phys. Rev. B}\ }\textbf {\bibinfo {volume}
  {92}},\ \bibinfo {pages} {235405} (\bibinfo {year} {2015})}\BibitemShut
  {NoStop}%
\bibitem [{\citenamefont {Makinistian}\ and\ \citenamefont
  {Albanesi}(2006)}]{lda-ges}%
  \BibitemOpen
  \bibfield  {author} {\bibinfo {author} {\bibfnamefont {L.}~\bibnamefont
  {Makinistian}}\ and\ \bibinfo {author} {\bibfnamefont {E.~A.}\ \bibnamefont
  {Albanesi}},\ }\href {\doibase 10.1103/PhysRevB.74.045206} {\bibfield
  {journal} {\bibinfo  {journal} {Phys. Rev. B}\ }\textbf {\bibinfo {volume}
  {74}},\ \bibinfo {pages} {045206} (\bibinfo {year} {2006})}\BibitemShut
  {NoStop}%
\bibitem [{\citenamefont {Cannuccia}\ and\ \citenamefont {Marini}(2011)}]{zpm}%
  \BibitemOpen
  \bibfield  {author} {\bibinfo {author} {\bibfnamefont {E.}~\bibnamefont
  {Cannuccia}}\ and\ \bibinfo {author} {\bibfnamefont {A.}~\bibnamefont
  {Marini}},\ }\href {\doibase 10.1103/PhysRevLett.107.255501} {\bibfield
  {journal} {\bibinfo  {journal} {Phys. Rev. Lett.}\ }\textbf {\bibinfo
  {volume} {107}},\ \bibinfo {pages} {255501} (\bibinfo {year}
  {2011})}\BibitemShut {NoStop}%
\bibitem [{\citenamefont {Giustino}\ \emph {et~al.}(2010)\citenamefont
  {Giustino}, \citenamefont {Louie},\ and\ \citenamefont {Cohen}}]{feliciano1}%
  \BibitemOpen
  \bibfield  {author} {\bibinfo {author} {\bibfnamefont {F.}~\bibnamefont
  {Giustino}}, \bibinfo {author} {\bibfnamefont {S.~G.}\ \bibnamefont {Louie}},
  \ and\ \bibinfo {author} {\bibfnamefont {M.~L.}\ \bibnamefont {Cohen}},\
  }\href {\doibase 10.1103/PhysRevLett.105.265501} {\bibfield  {journal}
  {\bibinfo  {journal} {Phys. Rev. Lett}\ }\textbf {\bibinfo {volume} {105}},\
  \bibinfo {pages} {265501} (\bibinfo {year} {2010})}\BibitemShut {NoStop}%
\bibitem [{\citenamefont {Kawai}\ \emph {et~al.}(2014)\citenamefont {Kawai},
  \citenamefont {Yamashita}, \citenamefont {Cannuccia},\ and\ \citenamefont
  {Marini}}]{hiroki2014}%
  \BibitemOpen
  \bibfield  {author} {\bibinfo {author} {\bibfnamefont {H.}~\bibnamefont
  {Kawai}}, \bibinfo {author} {\bibfnamefont {K.}~\bibnamefont {Yamashita}},
  \bibinfo {author} {\bibfnamefont {E.}~\bibnamefont {Cannuccia}}, \ and\
  \bibinfo {author} {\bibfnamefont {A.}~\bibnamefont {Marini}},\ }\href
  {\doibase 10.1103/PhysRevB.89.085202} {\bibfield  {journal} {\bibinfo
  {journal} {Phys. Rev. B}\ }\textbf {\bibinfo {volume} {89}},\ \bibinfo
  {pages} {085202} (\bibinfo {year} {2014})}\BibitemShut {NoStop}%
\bibitem [{\citenamefont {Marini}(2008)}]{marini2008}%
  \BibitemOpen
  \bibfield  {author} {\bibinfo {author} {\bibfnamefont {A.}~\bibnamefont
  {Marini}},\ }\href {\doibase 10.1103/PhysRevLett.101.106405} {\bibfield
  {journal} {\bibinfo  {journal} {Phys. Rev. Lett.}\ }\textbf {\bibinfo
  {volume} {101}},\ \bibinfo {pages} {106405} (\bibinfo {year}
  {2008})}\BibitemShut {NoStop}%
\bibitem [{\citenamefont {Molina-S\'{a}nchez}\ \emph
  {et~al.}(2016)\citenamefont {Molina-S\'{a}nchez}, \citenamefont {Palummo},
  \citenamefont {Marini},\ and\ \citenamefont {Wirtz}}]{alejandro2016}%
  \BibitemOpen
  \bibfield  {author} {\bibinfo {author} {\bibfnamefont {A.}~\bibnamefont
  {Molina-S\'{a}nchez}}, \bibinfo {author} {\bibfnamefont {M.}~\bibnamefont
  {Palummo}}, \bibinfo {author} {\bibfnamefont {A.}~\bibnamefont {Marini}}, \
  and\ \bibinfo {author} {\bibfnamefont {L.}~\bibnamefont {Wirtz}},\ }\href
  {\doibase 10.1103/PhysRevB.93.155435} {\bibfield  {journal} {\bibinfo
  {journal} {Phys. Rev. B}\ }\textbf {\bibinfo {volume} {93}},\ \bibinfo
  {pages} {155435} (\bibinfo {year} {2016})}\BibitemShut {NoStop}%
\bibitem [{\citenamefont {Zacharias}\ \emph {et~al.}(2015)\citenamefont
  {Zacharias}, \citenamefont {Patrick},\ and\ \citenamefont
  {Giustino}}]{feliciano0}%
  \BibitemOpen
  \bibfield  {author} {\bibinfo {author} {\bibfnamefont {M.}~\bibnamefont
  {Zacharias}}, \bibinfo {author} {\bibfnamefont {C.~E.}\ \bibnamefont
  {Patrick}}, \ and\ \bibinfo {author} {\bibfnamefont {F.}~\bibnamefont
  {Giustino}},\ }\href {\doibase 10.1103/PhysRevLett.115.177401} {\bibfield
  {journal} {\bibinfo  {journal} {Phys. Rev. Lett.}\ }\textbf {\bibinfo
  {volume} {115}},\ \bibinfo {pages} {177401} (\bibinfo {year}
  {2015})}\BibitemShut {NoStop}%
\bibitem [{\citenamefont {Ponc\'e}\ \emph {et~al.}(2015)\citenamefont
  {Ponc\'e}, \citenamefont {Gillet}, \citenamefont {Janssen}, \citenamefont
  {Marini}, \citenamefont {Verstraete},\ and\ \citenamefont {Gonze}}]{ponce}%
  \BibitemOpen
  \bibfield  {author} {\bibinfo {author} {\bibfnamefont {S.}~\bibnamefont
  {Ponc\'e}}, \bibinfo {author} {\bibfnamefont {Y.}~\bibnamefont {Gillet}},
  \bibinfo {author} {\bibfnamefont {J.~L.}\ \bibnamefont {Janssen}}, \bibinfo
  {author} {\bibfnamefont {A.}~\bibnamefont {Marini}}, \bibinfo {author}
  {\bibfnamefont {M.}~\bibnamefont {Verstraete}}, \ and\ \bibinfo {author}
  {\bibfnamefont {X.}~\bibnamefont {Gonze}},\ }\href {\doibase
  10.1063/1.4927081} {\bibfield  {journal} {\bibinfo  {journal} {J. Chem.
  Phys.}\ }\textbf {\bibinfo {volume} {143}},\ \bibinfo {pages} {102813}
  (\bibinfo {year} {2015})}\BibitemShut {NoStop}%
\bibitem [{\citenamefont {Cardona}\ and\ \citenamefont
  {Thewalt}(2005)}]{Cardonarev}%
  \BibitemOpen
  \bibfield  {author} {\bibinfo {author} {\bibfnamefont {M.}~\bibnamefont
  {Cardona}}\ and\ \bibinfo {author} {\bibfnamefont {M.~L.~W.}\ \bibnamefont
  {Thewalt}},\ }\href {\doibase 10.1103/RevModPhys.77.1173} {\bibfield
  {journal} {\bibinfo  {journal} {Rev. Mod. Phys.}\ }\textbf {\bibinfo {volume}
  {77}},\ \bibinfo {pages} {1173} (\bibinfo {year} {2005})}\BibitemShut
  {NoStop}%
\bibitem [{\citenamefont {Yu}\ and\ \citenamefont
  {Cardona}(2010)}]{cardonabook}%
  \BibitemOpen
  \bibfield  {author} {\bibinfo {author} {\bibfnamefont {P.~Y.}\ \bibnamefont
  {Yu}}\ and\ \bibinfo {author} {\bibfnamefont {M.}~\bibnamefont {Cardona}},\
  }\href@noop {} {\emph {\bibinfo {title} {Fundamentals of Semiconductors}}}\
  (\bibinfo  {publisher} {Springer-Verlag, Berlin},\ \bibinfo {year}
  {2010})\BibitemShut {NoStop}%
\bibitem [{\citenamefont {Shokhovets}\ \emph {et~al.}(2014)\citenamefont
  {Shokhovets}, \citenamefont {Barwolf}, \citenamefont {Gobsch}, \citenamefont
  {Runge}, \citenamefont {K{\"o}hler},\ and\ \citenamefont
  {Ambacher}}]{exci-ph}%
  \BibitemOpen
  \bibfield  {author} {\bibinfo {author} {\bibfnamefont {S.}~\bibnamefont
  {Shokhovets}}, \bibinfo {author} {\bibfnamefont {F.}~\bibnamefont {Barwolf}},
  \bibinfo {author} {\bibfnamefont {G.}~\bibnamefont {Gobsch}}, \bibinfo
  {author} {\bibfnamefont {E.}~\bibnamefont {Runge}}, \bibinfo {author}
  {\bibfnamefont {K.}~\bibnamefont {K{\"o}hler}}, \ and\ \bibinfo {author}
  {\bibfnamefont {O.}~\bibnamefont {Ambacher}},\ }\href {\doibase
  10.1002/pssc.201300311} {\bibfield  {journal} {\bibinfo  {journal} {physica
  status solidi (c)}\ }\textbf {\bibinfo {volume} {1}},\ \bibinfo {pages}
  {297–301} (\bibinfo {year} {2014})}\BibitemShut {NoStop}%
\bibitem [{\citenamefont {Giannozzi}\ and\ \citenamefont {\textit{et.
  al.}}(2009)}]{pwscf}%
  \BibitemOpen
  \bibfield  {author} {\bibinfo {author} {\bibfnamefont {P.}~\bibnamefont
  {Giannozzi}}\ and\ \bibinfo {author} {\bibnamefont {\textit{et. al.}}},\
  }\href@noop {} {\bibfield  {journal} {\bibinfo  {journal} {J. Phys. Condens.
  Matter}\ }\textbf {\bibinfo {volume} {21}},\ \bibinfo {pages} {395502}
  (\bibinfo {year} {2009})}\BibitemShut {NoStop}%
\bibitem [{\citenamefont {Baroni}\ \emph {et~al.}(2001)\citenamefont {Baroni},
  \citenamefont {de~Gironcoli}, \citenamefont {Corso},\ and\ \citenamefont
  {Giannozzi}}]{dfpt}%
  \BibitemOpen
  \bibfield  {author} {\bibinfo {author} {\bibfnamefont {S.}~\bibnamefont
  {Baroni}}, \bibinfo {author} {\bibfnamefont {S.}~\bibnamefont
  {de~Gironcoli}}, \bibinfo {author} {\bibfnamefont {A.~D.}\ \bibnamefont
  {Corso}}, \ and\ \bibinfo {author} {\bibfnamefont {P.}~\bibnamefont
  {Giannozzi}},\ }\href {\doibase 10.1103/RevModPhys.73.515} {\bibfield
  {journal} {\bibinfo  {journal} {Rev. Mod. Phys.}\ }\textbf {\bibinfo {volume}
  {73}},\ \bibinfo {pages} {515} (\bibinfo {year} {2001})}\BibitemShut
  {NoStop}%
\bibitem [{\citenamefont {Onida}\ \emph {et~al.}(2002)\citenamefont {Onida},
  \citenamefont {Reining},\ and\ \citenamefont {Rubio}}]{Onida2002}%
  \BibitemOpen
  \bibfield  {author} {\bibinfo {author} {\bibfnamefont {G.}~\bibnamefont
  {Onida}}, \bibinfo {author} {\bibfnamefont {L.}~\bibnamefont {Reining}}, \
  and\ \bibinfo {author} {\bibfnamefont {A.}~\bibnamefont {Rubio}},\ }\href
  {\doibase 10.1103/RevModPhys.74.601} {\bibfield  {journal} {\bibinfo
  {journal} {Rev. Mod. Phys.}\ }\textbf {\bibinfo {volume} {74}},\ \bibinfo
  {pages} {601} (\bibinfo {year} {2002})}\BibitemShut {NoStop}%
\bibitem [{\citenamefont {Ponc\'e}\ \emph
  {et~al.}(2014{\natexlab{a}})\citenamefont {Ponc\'e}, \citenamefont
  {Antonius}, \citenamefont {Gillet}, \citenamefont {Boulanger}, \citenamefont
  {Janssen}, \citenamefont {Marini}, \citenamefont {C\^ot\'e},\ and\
  \citenamefont {Gonze}}]{marini2014}%
  \BibitemOpen
  \bibfield  {author} {\bibinfo {author} {\bibfnamefont {S.}~\bibnamefont
  {Ponc\'e}}, \bibinfo {author} {\bibfnamefont {G.}~\bibnamefont {Antonius}},
  \bibinfo {author} {\bibfnamefont {Y.}~\bibnamefont {Gillet}}, \bibinfo
  {author} {\bibfnamefont {P.}~\bibnamefont {Boulanger}}, \bibinfo {author}
  {\bibfnamefont {J.~L.}\ \bibnamefont {Janssen}}, \bibinfo {author}
  {\bibfnamefont {A.}~\bibnamefont {Marini}}, \bibinfo {author} {\bibfnamefont
  {M.}~\bibnamefont {C\^ot\'e}}, \ and\ \bibinfo {author} {\bibfnamefont
  {X.}~\bibnamefont {Gonze}},\ }\href {\doibase 10.1103/PhysRevB.90.214304}
  {\bibfield  {journal} {\bibinfo  {journal} {Phys. Rev. B.}\ }\textbf
  {\bibinfo {volume} {90}},\ \bibinfo {pages} {214304} (\bibinfo {year}
  {2014}{\natexlab{a}})}\BibitemShut {NoStop}%
\bibitem [{\citenamefont {Marini}\ \emph {et~al.}(2015)\citenamefont {Marini},
  \citenamefont {ponc\'e},\ and\ \citenamefont {Gonze}}]{marini2015}%
  \BibitemOpen
  \bibfield  {author} {\bibinfo {author} {\bibfnamefont {A.}~\bibnamefont
  {Marini}}, \bibinfo {author} {\bibfnamefont {S.}~\bibnamefont {ponc\'e}}, \
  and\ \bibinfo {author} {\bibfnamefont {X.}~\bibnamefont {Gonze}},\ }\href
  {\doibase 10.1103/PhysRevB.91.224310} {\bibfield  {journal} {\bibinfo
  {journal} {Phys. Rev. B.}\ }\textbf {\bibinfo {volume} {91}},\ \bibinfo
  {pages} {224310} (\bibinfo {year} {2015})}\BibitemShut {NoStop}%
\bibitem [{\citenamefont {Ponc\'e}\ \emph
  {et~al.}(2014{\natexlab{b}})\citenamefont {Ponc\'e}, \citenamefont
  {Antonius}, \citenamefont {Boulanger}, \citenamefont {Cannuccia},
  \citenamefont {Marini}, \citenamefont {C\^ot\'e},\ and\ \citenamefont
  {Gonze}}]{q-random}%
  \BibitemOpen
  \bibfield  {author} {\bibinfo {author} {\bibfnamefont {S.}~\bibnamefont
  {Ponc\'e}}, \bibinfo {author} {\bibfnamefont {G.}~\bibnamefont {Antonius}},
  \bibinfo {author} {\bibfnamefont {P.}~\bibnamefont {Boulanger}}, \bibinfo
  {author} {\bibfnamefont {E.}~\bibnamefont {Cannuccia}}, \bibinfo {author}
  {\bibfnamefont {A.}~\bibnamefont {Marini}}, \bibinfo {author} {\bibfnamefont
  {M.}~\bibnamefont {C\^ot\'e}}, \ and\ \bibinfo {author} {\bibfnamefont
  {X.}~\bibnamefont {Gonze}},\ }\href {\doibase
  10.1016/j.commatsci.2013.11.031} {\bibfield  {journal} {\bibinfo  {journal}
  {physica status solidi (c)}\ }\textbf {\bibinfo {volume} {83}},\ \bibinfo
  {pages} {341–348} (\bibinfo {year} {2014}{\natexlab{b}})}\BibitemShut
  {NoStop}%
\bibitem [{\citenamefont {Marini}\ \emph {et~al.}(2009)\citenamefont {Marini},
  \citenamefont {Hogan}, \citenamefont {Gr{\"u}ning},\ and\ \citenamefont
  {Varsano}}]{Marini20091392}%
  \BibitemOpen
  \bibfield  {author} {\bibinfo {author} {\bibfnamefont {A.}~\bibnamefont
  {Marini}}, \bibinfo {author} {\bibfnamefont {C.}~\bibnamefont {Hogan}},
  \bibinfo {author} {\bibfnamefont {M.}~\bibnamefont {Gr{\"u}ning}}, \ and\
  \bibinfo {author} {\bibfnamefont {D.}~\bibnamefont {Varsano}},\ }\href@noop
  {} {\bibfield  {journal} {\bibinfo  {journal} {Computer Physics
  Communications}\ }\textbf {\bibinfo {volume} {180}},\ \bibinfo {pages} {1392}
  (\bibinfo {year} {2009})}\BibitemShut {NoStop}%
\bibitem [{\citenamefont {Cannuccia}\ and\ \citenamefont
  {Marini}(2012)}]{cannu}%
  \BibitemOpen
  \bibfield  {author} {\bibinfo {author} {\bibfnamefont {E.}~\bibnamefont
  {Cannuccia}}\ and\ \bibinfo {author} {\bibfnamefont {A.}~\bibnamefont
  {Marini}},\ }\href {\doibase 10.1103/PhysRevLett.107.255501} {\bibfield
  {journal} {\bibinfo  {journal} {Eur. Phys. Jour. B}\ }\textbf {\bibinfo
  {volume} {85}},\ \bibinfo {pages} {320} (\bibinfo {year} {2012})}\BibitemShut
  {NoStop}%
\bibitem [{\citenamefont {Gali}\ \emph {et~al.}(2015)\citenamefont {Gali},
  \citenamefont {Demj\'an}, \citenamefont {V{\"o}r{\"o}s}, \citenamefont
  {Thiering}, \citenamefont {Cannuccia},\ and\ \citenamefont {Marini}}]{gali}%
  \BibitemOpen
  \bibfield  {author} {\bibinfo {author} {\bibfnamefont {A.}~\bibnamefont
  {Gali}}, \bibinfo {author} {\bibfnamefont {T.}~\bibnamefont {Demj\'an}},
  \bibinfo {author} {\bibfnamefont {M.}~\bibnamefont {V{\"o}r{\"o}s}}, \bibinfo
  {author} {\bibfnamefont {G.}~\bibnamefont {Thiering}}, \bibinfo {author}
  {\bibfnamefont {E.}~\bibnamefont {Cannuccia}}, \ and\ \bibinfo {author}
  {\bibfnamefont {A.}~\bibnamefont {Marini}},\ }\href {\doibase
  10.1038/ncomms11327} {\bibfield  {journal} {\bibinfo  {journal} {Nat.
  Commun.}\ }\textbf {\bibinfo {volume} {7}},\ \bibinfo {pages} {11327}
  (\bibinfo {year} {2015})}\BibitemShut {NoStop}%
\bibitem [{\citenamefont {Bechstedt}\ \emph {et~al.}(1997)\citenamefont
  {Bechstedt}, \citenamefont {Tenelsen}, \citenamefont {Adolph},\ and\
  \citenamefont {Sole}}]{kernelstat1}%
  \BibitemOpen
  \bibfield  {author} {\bibinfo {author} {\bibfnamefont {F.}~\bibnamefont
  {Bechstedt}}, \bibinfo {author} {\bibfnamefont {K.}~\bibnamefont {Tenelsen}},
  \bibinfo {author} {\bibfnamefont {B.}~\bibnamefont {Adolph}}, \ and\ \bibinfo
  {author} {\bibfnamefont {R.~D.}\ \bibnamefont {Sole}},\ }\href {\doibase
  10.1103/PhysRevLett.78.1528} {\bibfield  {journal} {\bibinfo  {journal}
  {Phys. Rev. Lett.}\ }\textbf {\bibinfo {volume} {78}},\ \bibinfo {pages}
  {1528} (\bibinfo {year} {1997})}\BibitemShut {NoStop}%
\bibitem [{\citenamefont {Marini}\ and\ \citenamefont
  {Sole}(2003)}]{kernelstat2}%
  \BibitemOpen
  \bibfield  {author} {\bibinfo {author} {\bibfnamefont {A.}~\bibnamefont
  {Marini}}\ and\ \bibinfo {author} {\bibfnamefont {R.~D.}\ \bibnamefont
  {Sole}},\ }\href {\doibase 10.1103/PhysRevLett.91.176402} {\bibfield
  {journal} {\bibinfo  {journal} {Phys. Rev. Lett.}\ }\textbf {\bibinfo
  {volume} {91}},\ \bibinfo {pages} {176402} (\bibinfo {year}
  {2003})}\BibitemShut {NoStop}%
\bibitem [{\citenamefont {Marini}(2013)}]{ee1}%
  \BibitemOpen
  \bibfield  {author} {\bibinfo {author} {\bibfnamefont {A.}~\bibnamefont
  {Marini}},\ }\href {\doibase 10.1088/1742-6596/427/1/012003} {\bibfield
  {journal} {\bibinfo  {journal} {J. Phys.: Conf. Ser.}\ }\textbf {\bibinfo
  {volume} {427}},\ \bibinfo {pages} {012003} (\bibinfo {year}
  {2013})}\BibitemShut {NoStop}%
\bibitem [{\citenamefont {Marini}(2014)}]{ee2}%
  \BibitemOpen
  \bibfield  {author} {\bibinfo {author} {\bibfnamefont {A.}~\bibnamefont
  {Marini}},\ }\href {\doibase 10.1103/PhysRevLett.112.257402} {\bibfield
  {journal} {\bibinfo  {journal} {Phys. Rev. Lett.}\ }\textbf {\bibinfo
  {volume} {112}},\ \bibinfo {pages} {257402} (\bibinfo {year}
  {2014})}\BibitemShut {NoStop}%
\bibitem [{\citenamefont {Malone}\ and\ \citenamefont
  {Kaxiras}(2013)}]{gw-ges}%
  \BibitemOpen
  \bibfield  {author} {\bibinfo {author} {\bibfnamefont {B.~D.}\ \bibnamefont
  {Malone}}\ and\ \bibinfo {author} {\bibfnamefont {E.}~\bibnamefont
  {Kaxiras}},\ }\href {\doibase 10.1103/PhysRevB.87.245312} {\bibfield
  {journal} {\bibinfo  {journal} {Phys. Rev. B}\ }\textbf {\bibinfo {volume}
  {87}},\ \bibinfo {pages} {245312} (\bibinfo {year} {2013})}\BibitemShut
  {NoStop}%
\bibitem [{\citenamefont {Ling}\ \emph {et~al.}(2016)\citenamefont {Ling},
  \citenamefont {Huang}, \citenamefont {Hasdeo}, \citenamefont {Liang},
  \citenamefont {Parkin}, \citenamefont {Tatsumi}, \citenamefont {Puretzky},
  \citenamefont {Das}, \citenamefont {Sumpter}, \citenamefont {Geohegan},
  \citenamefont {Kong}, \citenamefont {Saito}, \citenamefont {Drndic},
  \citenamefont {Meunier},\ and\ \citenamefont {Dresselhaus}}]{dreselh}%
  \BibitemOpen
  \bibfield  {author} {\bibinfo {author} {\bibfnamefont {X.}~\bibnamefont
  {Ling}}, \bibinfo {author} {\bibfnamefont {S.}~\bibnamefont {Huang}},
  \bibinfo {author} {\bibfnamefont {E.~H.}\ \bibnamefont {Hasdeo}}, \bibinfo
  {author} {\bibfnamefont {L.}~\bibnamefont {Liang}}, \bibinfo {author}
  {\bibfnamefont {W.~M.}\ \bibnamefont {Parkin}}, \bibinfo {author}
  {\bibfnamefont {Y.}~\bibnamefont {Tatsumi}}, \bibinfo {author} {\bibfnamefont
  {A.~R. T. N. A.~A.}\ \bibnamefont {Puretzky}}, \bibinfo {author}
  {\bibfnamefont {P.~M.}\ \bibnamefont {Das}}, \bibinfo {author} {\bibfnamefont
  {B.~G.}\ \bibnamefont {Sumpter}}, \bibinfo {author} {\bibfnamefont {D.~B.}\
  \bibnamefont {Geohegan}}, \bibinfo {author} {\bibfnamefont {J.}~\bibnamefont
  {Kong}}, \bibinfo {author} {\bibfnamefont {R.}~\bibnamefont {Saito}},
  \bibinfo {author} {\bibfnamefont {M.}~\bibnamefont {Drndic}}, \bibinfo
  {author} {\bibfnamefont {V.}~\bibnamefont {Meunier}}, \ and\ \bibinfo
  {author} {\bibfnamefont {M.~S.}\ \bibnamefont {Dresselhaus}},\ }\href
  {\doibase 10.1021/acs.nanolett.5b04540} {\bibfield  {journal} {\bibinfo
  {journal} {Nano Lett.}\ }\textbf {\bibinfo {volume} {16}},\ \bibinfo {pages}
  {2260} (\bibinfo {year} {2016})}\BibitemShut {NoStop}%
\bibitem [{\citenamefont {Logothetidis}\ \emph {et~al.}(1985)\citenamefont
  {Logothetidis}, \citenamefont {Vi{\~n}a},\ and\ \citenamefont
  {Cardona}}]{cardonaGeS2}%
  \BibitemOpen
  \bibfield  {author} {\bibinfo {author} {\bibfnamefont {S.}~\bibnamefont
  {Logothetidis}}, \bibinfo {author} {\bibfnamefont {L.}~\bibnamefont
  {Vi{\~n}a}}, \ and\ \bibinfo {author} {\bibfnamefont {M.}~\bibnamefont
  {Cardona}},\ }\href {\doibase 10.1103/PhysRevB.31.2180} {\bibfield  {journal}
  {\bibinfo  {journal} {Phys. Rev. B}\ }\textbf {\bibinfo {volume} {31}},\
  \bibinfo {pages} {2180} (\bibinfo {year} {1985})}\BibitemShut {NoStop}%
\bibitem [{\citenamefont {Antonius}\ \emph {et~al.}(2014)\citenamefont
  {Antonius}, \citenamefont {Ponc\'e}, \citenamefont {Boulanger}, \citenamefont
  {C\^ot\'e},\ and\ \citenamefont {Gonze}}]{gonze}%
  \BibitemOpen
  \bibfield  {author} {\bibinfo {author} {\bibfnamefont {G.}~\bibnamefont
  {Antonius}}, \bibinfo {author} {\bibfnamefont {S.}~\bibnamefont {Ponc\'e}},
  \bibinfo {author} {\bibfnamefont {P.}~\bibnamefont {Boulanger}}, \bibinfo
  {author} {\bibfnamefont {M.}~\bibnamefont {C\^ot\'e}}, \ and\ \bibinfo
  {author} {\bibfnamefont {X.}~\bibnamefont {Gonze}},\ }\href {\doibase
  10.1103/PhysRevLett.112.215501} {\bibfield  {journal} {\bibinfo  {journal}
  {Phys. Rev. Lett.}\ }\textbf {\bibinfo {volume} {112}},\ \bibinfo {pages}
  {215501} (\bibinfo {year} {2014})}\BibitemShut {NoStop}%
\bibitem [{\citenamefont {Zhang}\ \emph {et~al.}(2016)\citenamefont {Zhang},
  \citenamefont {Wang}, \citenamefont {Liu}, \citenamefont {Huang},
  \citenamefont {Zhou}, \citenamefont {Cai}, \citenamefont {Xie}, \citenamefont
  {Yang}, \citenamefont {Chen},\ and\ \citenamefont {Zeng}}]{GeSmono}%
  \BibitemOpen
  \bibfield  {author} {\bibinfo {author} {\bibfnamefont {S.}~\bibnamefont
  {Zhang}}, \bibinfo {author} {\bibfnamefont {N.}~\bibnamefont {Wang}},
  \bibinfo {author} {\bibfnamefont {S.}~\bibnamefont {Liu}}, \bibinfo {author}
  {\bibfnamefont {S.}~\bibnamefont {Huang}}, \bibinfo {author} {\bibfnamefont
  {W.}~\bibnamefont {Zhou}}, \bibinfo {author} {\bibfnamefont {B.}~\bibnamefont
  {Cai}}, \bibinfo {author} {\bibfnamefont {M.}~\bibnamefont {Xie}}, \bibinfo
  {author} {\bibfnamefont {Q.}~\bibnamefont {Yang}}, \bibinfo {author}
  {\bibfnamefont {X.}~\bibnamefont {Chen}}, \ and\ \bibinfo {author}
  {\bibfnamefont {H.}~\bibnamefont {Zeng}},\ }\href {\doibase
  10.1088/0957-4484/27/27/274001} {\bibfield  {journal} {\bibinfo  {journal}
  {Nanotechnology}\ }\textbf {\bibinfo {volume} {27}},\ \bibinfo {pages}
  {274001} (\bibinfo {year} {2016})}\BibitemShut {NoStop}%
\end{thebibliography}

\end{document}